\documentclass[twocolumn,showpacs,preprintnumbers,amsmath,amssymb,superscriptaddress]{revtex4}
\usepackage{graphicx}
\usepackage{dcolumn}
\usepackage{bm}

\newcommand{\affd}[1]{Dipartimento di Fisica dell'Universit\`a e Sezione INFN, #1, Italy}
\renewcommand{\to}        {\ensuremath{\rightarrow}}
\def\ifm#1{\relax\ifmmode#1\else$#1$\fi}
\def\pt#1,#2,{\ifm{#1\x10^{#2}}}  \def\x{\ifm{\times}}
\def\ab{\ifm{\sim}}   \def\f{\ifm{\phi}} \def\pio{\ifm{\pi^0\pi^0}}
\def\pic{\ifm{\pi^+\pi^-}}   \def\gam{\ifm{\gamma}}
\def\epm{\ifm{e^+e^-}} 
\let\cl=\centerline  
\def\figbox#1;#2;{\parbox{#2cm}{\epsfig{file=#1.eps,width=#2cm}}}
\def\figboxc#1;#2;{\cl{\figbox#1;#2;}}
\makeatletter
\newdimen\z@ \z@=0pt 
\newskip\z@skip \z@skip=0pt plus0pt minus0pt
\def\m@th{\mathsurround=\z@}
\def\ialign{\everycr{}\tabskip\z@skip\halign} 
\def\eqalign#1{\null\,\vcenter{\openup\jot\m@th
  \ialign{\strut\hfil$\displaystyle{##}$&$\displaystyle{{}##}$\hfil
    \crcr#1\crcr}}\,}
\makeatother

\newcommand{\R}{\ensuremath{R_{S}^{\pi}}}
\newcommand{\RL}{\ensuremath{R_{L}^{\pi}}}
\newcommand{\Dafne}       {DA\char8NE}
\newcommand{\kzero}       {\ensuremath{K^{0}}}
\newcommand{\kzerob}      {\ensuremath{\bar{K}^{0}}}
\newcommand{\ks}          {\ensuremath{K_{S}}}
\newcommand{\kl}          {\ensuremath{K_{L}}}

\newcommand{\Ppio}{\ensuremath{\pi^0}}

\newcommand{\tzero}{\ensuremath{T_{0}}}
\newcommand{\tzeroev}{\ensuremath{\tzero}}
\newcommand{\tzeromc}{\ensuremath{\tzero^\mathrm{true}}}
\newcommand{\tbunch}{\ensuremath{T_\mathrm{RF}}}
\newcommand{\pin}{\ensuremath{P^{\rm \,in}_{n}}}
\newcommand{\pout}{\ensuremath{P^{\rm\,out}_{n}}}
\newcommand{\tn}{\ensuremath{f_{n}}}

\newcommand{\bstar}{\ensuremath{\beta^{\ast}}}

\newcommand{\Racc}{\ensuremath{R_{\rm acc}}}
\newcommand{\phikskl}{\ensuremath{\phi\rightarrow K_S K_L}}

\newcommand{\DKSeIII}{\ensuremath{K_S\rightarrow\pi e \nu}}
\newcommand{\DKSmuIII}{\ensuremath{K_S\rightarrow\pi\mu\nu}}
\newcommand{\DKSeIIIboth}{\ensuremath{K_S\rightarrow\pi^{\mp}e^{\pm}\nu(\overline{\nu})}}
\newcommand{\DKSeIIIeppm}{\ensuremath{K_S\rightarrow\pi^{-}e^{+}\nu}}
\newcommand{\DKSeIIIempp}{\ensuremath{K_S\rightarrow\pi^{+}e^{-}\overline{\nu}}}
\newcommand{\DKSpippim}{\ensuremath{K_S\rightarrow\pi^+\pi^-(\gamma)}}
\newcommand{\KSpippim}{\ensuremath{K_S\rightarrow\pi^+\pi^-}}
\newcommand{\pippim}{\ensuremath{\pi^+\pi^-}}
\newcommand{\DKSpiopio}{\ensuremath{K_S\rightarrow\pi^0\pi^0}}

\newcommand{\piopio}{\ensuremath{\pi^0\pi^0}}

\newcommand{\DKLpiopiopio}{\ensuremath{K_L\rightarrow\pi^0\pi^0\pi^0}}
\newcommand{\DKLpippimpio}{\ensuremath{K_L\rightarrow\pi^+\pi^-\pi^0}}

\newcommand{\Dphipippimpio}{\ensuremath{\phi\rightarrow\pi^+\pi^-\pi^0}}
\newcommand{\eV}{{e\kern-.07em V}}
\newcommand{\MeV}{\,{\rm M\eV}}
\newcommand{\MeVsuc}{\,{\rm M\eV\ensuremath{/c}}}
\newcommand{\GeV}{{\rm \,G\eV}}
\newcommand{\keV}{{\rm \,k\eV}}

\newcommand{\nb}{{\rm \,nb}}
\newcommand{\ps}{{\rm \,ps}}
\newcommand{\ns}{{\rm \,ns}}
\newcommand{\mm}{{\rm \,mm}}
\newcommand{\m}{{\rm \,m}}
\newcommand{\cm}{{\rm \,cm}}
\newcommand{\um}{\,\textrm{$\mu$m}}
\newcommand{\T}{{\rm \,T}}
\newcommand{\Lpb}{\ensuremath{\rm \,pb^{-1}}}
\newcommand{\vecbf}[1]{\ensuremath{\mathbf{#1}}}
\newcommand{\BR}[1]{\ensuremath{\mathrm{BR}(#1)}}
\newcommand{\gammo}[1]{\ensuremath{\Gamma(#1)}}

\newcommand{\rT}{\ensuremath{\rho_L}}
\newcommand{\kcr}{\ensuremath{K_\mathrm{cr}}}
\newcommand{\kcrmc}{\ensuremath{K_\mathrm{cr}^\mathrm{true}}}
\newcommand{\ecr}{\ensuremath{E_{\rm cr}}}
\newcommand{\kk}{\ensuremath{\rm cr}}
\newcommand{\nk}{\ensuremath{\overline{\rm cr}}}
\newcommand{\ek}{\ensuremath{\epsilon_{\rm cr}}}
\newcommand{\ecl}{\ensuremath{\epsilon_{\mathrm{cl}}}}
\newcommand{\etrk}{\ensuremath{\epsilon_{\mathrm{trk}}}}
\newcommand{\eto}{\ensuremath{\epsilon_{\mathrm{T0}}}}
\newcommand{\pspli}{\ensuremath{P_{\rm split}}}

\newcommand{\dmin}{\ensuremath{d_{\rm min}}}
\newcommand{\SN}[2]{\ensuremath{#1\times10^{#2}}}
\newcommand{\VS}[2]{\ensuremath{#1\pm#2}}

\newcommand{\Fig}{Fig.\,}
\newcommand{\Ref}{Ref.\,}
\newcommand{\Eq}{Eq.\,}
\newcommand{\Tab}{Tab.\,}

\newcommand{\Virginia}{Physics Department, University of Virginia, Charlottesville, VA, USA}
\newcommand{\Frascati}{Laboratori Nazionali di Frascati dell'INFN, Frascati, Italy}
\newcommand{\Karlsruhe}{Institut f\"ur Experimentelle Kernphysik, Universit\"at Karlsruhe, Germany}
\newcommand{\Lecce}{\affd{Lecce}}
\newcommand{\Na}{Dipartimento di Scienze Fisiche dell'Universit\`a ``Federico II'' e Sezione INFN, Napoli, Italy}
\newcommand{\Energ}{Dipartimento di Energetica dell'Universit\`a ``La Sapienza'', Roma, Italy}
\newcommand{\Romaone}{\affd{``La Sapienza''}}
\newcommand{\Romatwo}{\affd{``Tor Vergata''}}
\newcommand{\Romathree}{\affd{``Roma Tre''}}
\newcommand{\Pisa}{\affd{Pisa}}
\newcommand{\StonyBrook}{Physics Department, State University of New York at Stony Brook, NY, USA}
\newcommand{\Beijing}{Permanent address: Institute of High Energy Physics, CAS, Beijing, China}
\newcommand{\Moscow}{Permanent address: Institute for Theoretical and Experimental Physics, Moscow, Russia}

\begin{document}
\title{\mathversion{bold}Precise measurement of $\gammo{\DKSpippim}/\gammo{\DKSpiopio}$ with the KLOE detector at \Dafne}
\author{F.~Ambrosino}\affiliation{\Na}
\author{A.~Antonelli}\affiliation{\Frascati}
\author{M.~Antonelli}\affiliation{\Frascati}
\author{C.~Bacci}\affiliation{\Romathree}
\author{P.~Beltrame}\affiliation{\Karlsruhe}
\author{G.~Bencivenni}\affiliation{\Frascati}
\author{S.~Bertolucci}\affiliation{\Frascati}
\author{C.~Bini}\affiliation{\Romaone}
\author{C.~Bloise}\affiliation{\Frascati}
\author{V.~Bocci}\affiliation{\Romaone}
\author{F.~Bossi}\affiliation{\Frascati}
\author{D.~Bowring}\affiliation{\Virginia}
\author{P.~Branchini}\affiliation{\Romathree}
\author{R.~Caloi}\affiliation{\Romaone}
\author{P.~Campana}\affiliation{\Frascati}
\author{G.~Capon}\affiliation{\Frascati}
\author{T.~Capussela}\affiliation{\Na}                
\author{F.~Ceradini}\affiliation{\Romathree}             
\author{S.~Chi}\affiliation{\Frascati}          
\author{G.~Chiefari}\affiliation{\Na}                
\author{P.~Ciambrone}\affiliation{\Frascati}          
\author{S.~Conetti}\affiliation{\Virginia}          
\author{E.~De~Lucia}\affiliation{\Frascati}          
\author{A.~De~Santis}\affiliation{\Romaone}             
\author{P.~De~Simone}\affiliation{\Frascati}          
\author{G.~De~Zorzi}\affiliation{\Romaone}             
\author{S.~Dell'Agnello}\affiliation{\Frascati}          
\author{A.~Denig}\affiliation{\Karlsruhe}         
\author{A.~Di~Domenico}\affiliation{\Romaone}             
\author{C.~Di~Donato}\affiliation{\Na}                
\author{S.~Di~Falco}\affiliation{\Pisa}              
\author{B.~Di~Micco}\affiliation{\Romathree}             
\author{A.~Doria}\affiliation{\Na}                
\author{M.~Dreucci}\affiliation{\Frascati}          
\author{G.~Felici}\affiliation{\Frascati}          
\author{A.~Ferrari}\affiliation{\Frascati}         
\author{M.~L.~Ferrer}\affiliation{\Frascati}          
\author{G.~Finocchiaro}\affiliation{\Frascati}          
\author{S.~Fiore}\affiliation{\Romaone}             
\author{C.~Forti}\affiliation{\Frascati}          
\author{P.~Franzini}\affiliation{\Romaone}             
\author{C.~Gatti}\affiliation{\Frascati}\email[Corresponding author: Claudio Gatti
INFN - LNF, Casella postale 13, 00044 Frascati (Roma), 
Italy; tel. +39-06-94032727]{claudio.gatti@lnf.infn.it}
\author{P.~Gauzzi}\affiliation{\Romaone}             
\author{S.~Giovannella}\affiliation{\Frascati}          
\author{E.~Gorini}\affiliation{\Lecce}             
\author{E.~Graziani}\affiliation{\Romathree}             
\author{M.~Incagli}\affiliation{\Pisa}              
\author{W.~Kluge}\affiliation{\Karlsruhe}         
\author{V.~Kulikov}\affiliation{\Moscow}            
\author{F.~Lacava}\affiliation{\Romaone}             
\author{G.~Lanfranchi}\affiliation{\Frascati}          
\author{J.~Lee-Franzini}\affiliation{\Frascati}\affiliation{\StonyBrook}
\author{D.~Leone}\affiliation{\Karlsruhe}         
\author{M.~Martini}\affiliation{\Frascati}          
\author{P.~Massarotti}\affiliation{\Na}                
\author{W.~Mei}\affiliation{\Frascati}          
\author{S.~Meola}\affiliation{\Na}                
\author{S.~Miscetti}\affiliation{\Frascati}          
\author{M.~Moulson}\affiliation{\Frascati}          
\author{S.~M\"uller}\affiliation{\Frascati}
\author{F.~Murtas}\affiliation{\Frascati}          
\author{M.~Napolitano}\affiliation{\Na}                
\author{F.~Nguyen}\affiliation{\Romathree}             
\author{M.~Palutan}\affiliation{\Frascati}\email[Corresponding author: Matteo Palutan
INFN - LNF, Casella postale 13, 00044 Frascati (Roma), 
Italy; tel. +39-06-94032697]{matteo.palutan@lnf.infn.it}          
\author{E.~Pasqualucci}\affiliation{\Romaone}             
\author{A.~Passeri}\affiliation{\Romathree}             
\author{V.~Patera}\affiliation{\Frascati}\affiliation{\Energ}
\author{F.~Perfetto}\affiliation{\Na}                
\author{L.~Pontecorvo}\affiliation{\Romaone}             
\author{M.~Primavera}\affiliation{\Lecce}             
\author{P.~Santangelo}\affiliation{\Frascati}          
\author{E.~Santovetti}\affiliation{\Romatwo}             
\author{G.~Saracino}\affiliation{\Na}                
\author{B.~Sciascia}\affiliation{\Frascati}          
\author{A.~Sciubba}\affiliation{\Frascati}\affiliation{\Energ}    
\author{F.~Scuri}\affiliation{\Pisa}              
\author{I.~Sfiligoi}\affiliation{\Frascati}          
\author{T.~Spadaro}\affiliation{\Frascati}\email[Corresponding author: Tommaso Spadaro
INFN - LNF, Casella postale 13, 00044 Frascati (Roma), 
Italy; tel. +39-06-94032698]{tommaso.spadaro@lnf.infn.it}
\author{M.~Testa}\affiliation{\Romaone}             
\author{L.~Tortora}\affiliation{\Romathree}             
\author{P.~Valente}\affiliation{\Frascati}          
\author{B.~Valeriani}\affiliation{\Karlsruhe}         
\author{G.~Venanzoni}\affiliation{\Frascati}          
\author{S.~Veneziano}\affiliation{\Romaone}             
\author{A.~Ventura}\affiliation{\Lecce}             
\author{R.~Versaci}\affiliation{\Frascati}         
\author{G.~Xu}\affiliation{\Frascati}\affiliation{\Beijing}

\collaboration{The KLOE Collaboration}\noaffiliation
\begin{abstract}
Using a sample of over 400 million $\phi\!\to\!\ks\kl$ decays produced during the years 2001 and 2002
at the \Dafne\ $e^+e^-$ collider, 
the ratio $\R=\gammo{\DKSpippim}/\gammo{\DKSpiopio}$ has been measured with the KLOE detector. 
The result is $\R=2.2555\pm 0.0012_{\rm stat}\pm 0.0021_{\rm syst\mbox{-}stat}\pm 0.0050_{\rm syst}$,
which is in good agreement with the previously published result based on the KLOE data sample from the year 2000. 
The average of the KLOE results is 
$\R=2.2549 \pm 0.0054$, reducing the total error by a factor of three, to $0.25\%$.
\end{abstract}
\pacs{13.25.Es, 14.40.Aq}

\maketitle

\section{Introduction}
\label{intro}
The ratio \R\,=\,\gammo{\DKSpippim}/\gammo{\DKSpiopio} is a fundamental parameter of the \ks\ meson. 
Since the sum of the branching ratios (BR's) for the two dominant decays of the 
short-lived neutral kaon differs from unity by just $10^{-3}$, 
the measurement of \R\ provides the BR's for \DKSpiopio\ and \DKSpippim\ with only small corrections. 
The latter BR is a convenient normalization for 
the BR's of all other \ks\ decays to charged particles. In particular, it is used to obtain  \gammo{\DKSeIII}, 
which is of interest in testing many predictions of the Standard Model, as discussed in 
\Ref\onlinecite{plbpennew}. From \R\ one can also derive phenomenological parameters of the kaon system such as 
the relative magnitude and phase of the $I\!=\!0$ and $I\!=\!2$ $\pi\pi$-scattering amplitudes. Isospin-breaking 
effects and radiative corrections to the scattering amplitudes are discussed in Refs.~\onlinecite{cirigliano00,cirigliano04}.
Finally, \R\ enters into the double ratio that quantifies  
direct $CP$ violation in $K\to\pi\pi$ transitions:
\begin{equation}
\label{epsilonprimo}
\R/\RL=1 - 6\Re(\epsilon^\prime/\epsilon),
\end{equation}
where \RL\,=\,\gammo{\kl\to\pic(\gam)}/\gammo{\kl\to\pio}. The most accurate measurement of \R\ to date was performed
by KLOE using data collected in 2000 for an integrated luminosity of \ab17\Lpb:
 $\R\!=\!2.236 \pm 0.003_{\rm stat} \pm 0.015_{\rm syst}$~\cite{plb_rappo}. This result, which was more precise than 
the PDG average at the time~\cite{Groom:2000in}, for the first time properly included photon radiation and
increased the PDG value for \BR{\DKSpippim} by 0.5\%~\cite{PDBook}. The overall accuracy of the previous result, 0.7\%, 
was limited by systematic uncertainties. The present result is based on the analysis of 410\Lpb\ of integrated 
luminosity acquired during the
years 2001 and 2002, and improves on the total error by a factor of three, to $0.25\%$.

The paper is organized as follows. In the next section, a brief description of the KLOE detector
is given. In Section~\ref{scheme}, the selection criteria for the decays of interest are summarized.
In Section~\ref{effi}, a general description of the scheme used to evaluate the efficiency corrections
is given, followed by a detailed discussion on the tagging efficiencies, acceptances, and trigger efficiencies.
The result of the analysis is presented in Section~\ref{result}.

\section{Experimental setup}
\label{setup}
The data were collected with the KLOE detector at \Dafne, the Frascati \f\ factory. \Dafne\ 
is an \epm\ collider that operates at a center-of-mass energy of \ab1020\MeV, the mass of the \f\ meson. 
Positron and electron beams of equal energy collide at an angle of $\pi-25$\,mrad, producing \f\ mesons with a
small momentum in the horizontal plane: $p_{\phi}\ab13$\MeVsuc.
\f\ mesons decay \ab34\% of the time into nearly collinear \kzero\kzerob\ pairs. Because 
$J^{PC}(\f)\!=\!1^{--},$ the kaon pair is in an antisymmetric state, so that the final state is
 always \ks\kl. The contamination from $\kl\,\kl$ and $\ks\,\ks$ final states
is negligible for the purposes of this measurement~\cite{Dunietz:1987jf,Paver:1990fn,Close:1992ay}. 
Therefore, the detection of a \kl\ signals the presence of a \ks\ of known momentum and direction,
 independently of its decay mode.
This technique is called \textit{\ks\ tagging} in the following.
A total of \ab1.3 billion \f\ mesons were produced, yielding \ab430 million $\ks\,\kl$ pairs.

\begin{figure}[ht]
\begin{center}
    \mbox{
      \resizebox{!}{8cm}{\includegraphics{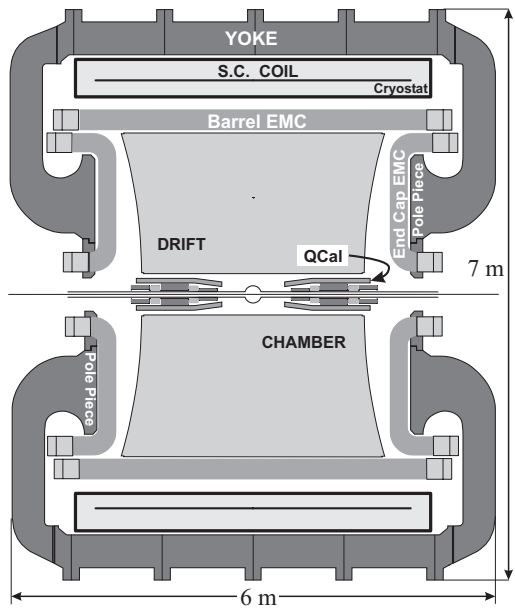}}
     }
\caption{Vertical cross section of the KLOE detector.}
\label{fig:kloedetector}
 \end{center}
\end{figure}
The KLOE detector 
(\Fig\ref{fig:kloedetector})
consists of a large cylindrical drift chamber (DC) surrounded by a 
lead/scintillating-fiber sampling calorimeter (EMC). A superconducting coil
surrounding the calorimeter provides a 0.52\T\ magnetic field. 
The drift chamber~\cite{DCnim}, which is 4\m\ in diameter and 3.3\m\ long, 
has 12,582 all-stereo tungsten 
sense wires and 37,746 aluminum field wires.
The chamber shell is made of carbon-fiber/epoxy composite, and the gas used is a 
90\% helium, 10\% isobutane mixture. 
These features maximize transparency to photons and reduce $\kl\!\to\!\ks$ regeneration and 
multiple scattering. 
The DC position resolutions are $\sigma_{xy}\approx 150\um$ 
and $\sigma_z\approx 2\mm.$ The momentum resolution is 
$\sigma(p_{\perp})/p_{\perp}\approx 0.4\%.$ Vertices are 
reconstructed with a spatial resolution of $\ab3\mm.$
The amount of material traversed before particles enter the DC volume 
affects the detection efficiency for \ks\ decay products.
Particles traverse the beam pipe and the inner DC wall, which are made
of a 500\um-thick layer of Albemet alloy (60\%\,Al-40\%\,Be) and a 800\um-thick layer of carbon-fiber/epoxy composite
aluminized on each side with a foil of 100\um.
The total amount of material corresponds to $\ab0.5\%\,{\rm X_0}$ and to an average
conversion probability of $\ab0.4\%$ for each photon from a \DKSpiopio\ decay.
Moreover, assuming a disappearance (including absorption, charge exchange, and inelastic processes) 
cross section of 400~mb for $\pi^{\pm}$ with $p\!=\!200$\MeVsuc\ interacting on 
carbon~\cite{navon}, and
using the same value for beryllium and aluminum,
the average probability of disappearance for each pion emitted from a \DKSpippim\ decay is $\ab0.5\%$.

The calorimeter~\cite{EmCnim} is divided into a barrel and two endcaps, contains a 
total of 88 modules, and covers 98\% of the solid angle. 
The modules are read out at both ends by photomultiplier tubes. 
The arrival times of particles and the three-dimensional positions of 
the energy deposits are determined from the signals at the two ends. 
The readout granularity is 
$\ab4.4\!\times\!4.4\cm^2$; the 2440 ``cells'' are arranged in five layers.
Cells close in time and space are grouped into a ``calorimeter cluster.''
For each cluster, the energy $E_{\rm cl}$ is the sum of the cell energies, and
the time $t_{\mathrm{cl}}$ and position $\vecbf{r}_{\mathrm{cl}}$ are calculated as
energy-weighted averages over the fired cells. The energy and time resolutions 
are $\sigma_E/E \!=\! 5.7\%/\!\sqrt{E (\GeV)}$ and  
$\sigma_t\!=\!57\ps/\!\sqrt{E (\GeV)}\oplus100\ps,$ respectively.

 Only the 
calorimeter trigger~\cite{TRGnim} is used for the present measurement.  
This requires two local energy deposits 
(trigger {\it sectors}) above a threshold of
$50$\MeV\ in the barrel and $150$\MeV\ in the endcaps. Events with only two fired trigger sectors in the
same endcap are rejected, because this topology is dominated by machine background. 
A single particle hitting the calorimeter barrel and releasing enough energy to fire two contiguous 
sectors generates a valid trigger.

Recognition and rejection of cosmic-ray events is also performed at the trigger level:
events with two energy deposits above a 30\MeV\ threshold in the outermost 
calorimeter plane are rejected as cosmic-ray events.
Moreover,
to reject residual cosmic rays and machine background events an offline software filter (FILFO) 
exploits calorimeter and DC information before tracks are reconstructed~\cite{OFFnim}.

The trigger has a large time spread with respect to the beam crossing time. However, it is 
synchronized with the machine RF divided by 4, $T_{\rm sync}\!\ab10.8\ns,$ 
with an accuracy of 50\ps. 
The time of the bunch crossing producing an event is determined offline during event reconstruction. 

The response of the detector to the decays of interest and the various backgrounds were studied by using
 the KLOE Monte Carlo
 (MC) simulation program~\cite{OFFnim}. 
Changes in the machine operation and background conditions are simulated on a 
run-by-run basis to improve agreement with data when averaged over the sample. 
The most important parameters are the beam energies and the crossing angle, which are obtained from the
analysis of Bhabha scattering events with $e^\pm$ polar angles above 45 degrees.
The average value of the center-of-mass energy is evaluated with a precision of 
30\keV\ for each $100\nb^{-1}$ of integrated luminosity.   

Particularly important for correct evaluation of the acceptance for \pippim\ and \piopio\
events is the rate of accidental clusters from the machine (\Racc). This is 
extracted from the analysis of $\epm\!\to\!\gamma\gamma$ events, where the low-energy and out-of-time
hits due to machine background are easily separated from the two 510\MeV\ photon clusters.

For the present analysis,
an MC sample of \phikskl\ decays that corresponds to an integrated luminosity of \ab550\Lpb\ is used;
for the other \f-meson final states, an MC sample equivalent
to \ab90\Lpb\ of integrated luminosity has been used.  
\section{Signal selection}
\label{scheme}
The mean decay lengths of the \ks\ and \kl\ are $\lambda_{S}\ab0.6\cm$ and
$\lambda_{L}\ab350\cm$, respectively. About 50\% of \kl's therefore reach the calorimeter before decaying. 
The \kl\ interaction in the calorimeter barrel (\kl\ {\it crash}, \kcr\ in the following) is identified 
by requiring a cluster of energy above a given threshold \ecr\
not associated with any track, and whose time corresponds to a velocity 
$\beta\!=\!r_{\mathrm{cl}}/c\,t_{\mathrm{cl}}$ 
compatible with the kaon velocity in the \f\ center of mass, $\bstar\!\ab0.216$, after the 
residual \f\ motion is considered. Events with clusters with $0.17\!\leq\!\bstar\!\leq\!0.28$ are selected.
These \kcr\ events are used to tag a \ks\ ``beam'' of known momentum.
The \ks\ trajectory is determined with an angular resolution of $1^\circ$ and 
the \ks\ momentum is evaluated with a resolution better than
2\MeVsuc\ from $\vecbf{p}_{\ks}\!=\!\vecbf{p}_{\phi}-\vecbf{p}_{\kl}$, where
the \kl\ momentum $\vecbf{p}_{\kl}$ is calculated by using 
the values of the center-of-mass energy and of the $\phi$ momentum $\vecbf{p}_{\phi},$
and the position of the \kcr\ cluster.

The interaction time, which must be known for the measurement of the cluster
times, is obtained from the first particle reaching the calorimeter (pions or photons for 
the events of interest) assuming a velocity $\beta\!=\!1.$ 
This definition of 
interaction time (\tzeroev\ in the following) does not require the \ks\ decay to be identified
when applying the tagging algorithm.
In order to reduce the probability that \tzeroev\ is accidentally determined from 
a particle due to machine background, 
the \tzeroev\ is required to be given by a cluster with 
energy $E_{\rm cl}\!>\!50$\MeV\ and distance to the beam line $\rho_{\rm cl}\!>\!60\cm.$
This is referred to as a ``\tzero\ cluster.''

\DKSpippim\ events are selected by requiring the presence of
two tracks of opposite charge with their point of closest approach to the origin inside a 
cylinder 4\cm\ in radius and 10\cm\ in length
along the beam line. 
The tracks momenta and polar angles must satisfy the fiducial cuts 
$120\!\leq\!p\!\leq\!300$\MeVsuc\ and  
$30^\circ\!\leq\!\theta\!\leq\!150^\circ.$ 
The tracks must also reach the EMC without spiralling, 
and at least one of them must have an associated \tzero\ cluster.

\DKSpiopio\ events are identified by the prompt photon clusters from $\pi^{0}$ decays. 
A prompt photon cluster must satisfy 
$|t_{\mathrm{cl}}-r_{\mathrm{cl}}/c|\!\le\!5\sigma_t,$
$\sigma_t$ being the energy-dependent time resolution, and must not be associated 
to any track. Machine background is reduced by cuts on the cluster energy and polar angle:
$E_{\rm cl}\!>\!20$\MeV\ and $|\cos\theta|\!<\!0.9.$
To accept a \DKSpiopio\ event, three or more prompt photons are required.

The numbers $N$ of \pippim\ and \piopio\ events and the corresponding selection efficiencies 
$\epsilon_{\rm sel}$ are then used to compute \R:
\begin{equation}
\label{strategone}
\R =
\frac{N(\pippim)}{N(\piopio)}
\frac{\epsilon_{\rm sel}(\piopio)}{\epsilon_{\rm sel}(\pippim)} 
\frac{C(\pippim)}{C(\piopio)},
\end{equation} 
where $C$ is the purity of the sample (the fraction of selected events that are signal), as evaluated 
from MC.

\section{Efficiency evaluation}
\label{effi}
\subsection{General scheme}
\label{effischeme}
The fractional statistical error from the counting is $\ab0.5\times10^{-3}$; 
the overall uncertainty is dominated by systematics. Therefore, in the analysis,
great effort has been put into carefully estimating all possible systematic effects, as discussed in detail
in \Ref\onlinecite{notappoonew}. 

The selection efficiency is expressed
for each of the two channels (\pippim, \piopio) as follows:
\begin{equation}
  \label{accesimple}
  \epsilon_{\rm sel}=\epsilon_{\rm tag + acc}\,\epsilon_{\rm trg}\,\epsilon_{\rm CV}\,\epsilon_{\rm FILFO},
\end{equation}
where
$\epsilon_{\rm tag + acc}$ is the joint efficiency for reconstructing both the tagging \kl\ interaction
and the \ks\ decay;
$\epsilon_{\rm trg}$, 
$\epsilon_{\rm CV}$, and $\epsilon_{\rm FILFO}$ are
the efficiencies for the trigger, the cosmic-ray veto, and the offline 
background filter (FILFO).
The tagging efficiency and the signal acceptance are correlated by the \tzeroev\ determination, 
and cannot be simply factorized, as discussed below.

\begin{figure}[ht]
\begin{center}
    \mbox{
     \resizebox{!}{6cm}{\includegraphics{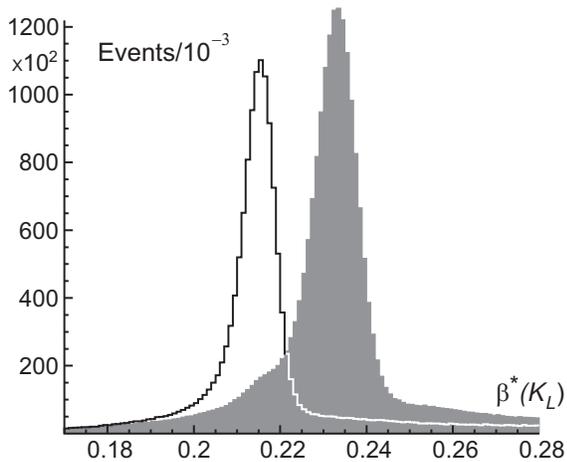}}
     }
\caption{\kl\ velocity transformed to the \f\ rest frame, \bstar, for \DKSpiopio\ (open histogram) and 
\KSpippim\ (shaded histogram). The range shown corresponds to the accepted window in \bstar.}
\label{betaspectrum}
 \end{center}
\end{figure}

For essentially all of selected \DKSpiopio\ events, the \tzeroev\ corresponds 
to the true collision time: if this is not the case, the prompt photon cluster 
selection fails and the event is lost.
For the purposes of \kcr\ selection, the velocity \bstar\ is therefore correctly evaluated
(open histogram of \Fig\ref{betaspectrum}).
In contrast, for most \KSpippim\ events, the \tzeroev\ does not correspond to the true collision time. 
Most charged pions arrive at the EMC $\ab3\ns$ 
later than \gam's from $\pi^0$ decays and the time \tzeroev\ is therefore
delayed by one RF period, $\tbunch \ab2.7\ns$; in a few percent of the cases, larger displacements 
are observed.
This results in a \ab10\% overestimation of the \kl\ velocity and a difference in the tail populations
(shaded histogram of \Fig\ref{betaspectrum}). 
The displacement of the two \bstar\ distributions within 
the accepted \bstar\ region affects the tagging efficiency, which then differs for each of the two
final states.  
In order to parametrize this effect, two classes of events are defined:   
events selected 
by the \kcr\ algorithm on the basis of the true value of the collision time are called \kcrmc,
the rejected ones are non-\kcrmc. 
While all of \DKSpiopio\ events are \kcrmc, 
the net effect due to incorrect \tzeroev\ determination on the tagging efficiency for \KSpippim\ events is that
a fraction $(1-A)\ab3\%$ of \kcrmc\ is lost, while a fraction $B\ab0.3\%$ 
of non-\kcrmc\ creeps into the selection (the fractions $A$ and $B$ are defined more precisely in 
Sec.~\ref{tag}).

\begin{figure}[ht]
\begin{center}
    \mbox{
     \resizebox{!}{6cm}{\includegraphics{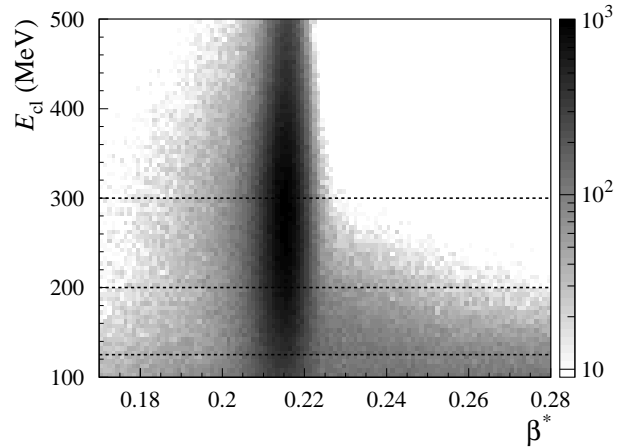}}
     }
\caption{\kcr\ cluster energy versus \bstar\ for \DKSpiopio\ events. 
The dashed lines correspond to the three different cuts on \kcr\ energy used in the analysis.}
\label{fig:evsbeta}
 \end{center}
\end{figure}

\begin{figure}[ht]
\begin{center}
    \mbox{
     \resizebox{!}{6cm}{\includegraphics{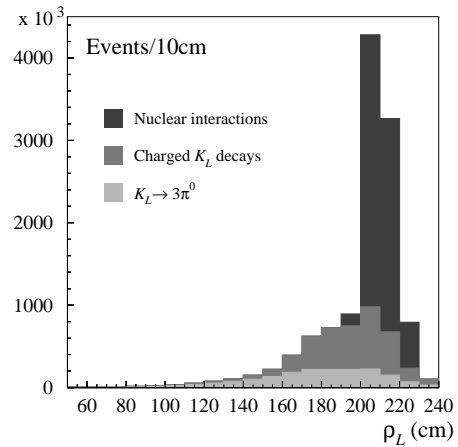}}
     }
\caption{MC distribution of \kl\ decay/interaction position \rT\ for \kcrmc\ events selected with 
$\ecr\!=\!125$\MeV; contributions
from \DKLpiopiopio, from \kl\ decays to charged particles, and from nuclear interactions are shown separately.}
\label{fig:rtspectra}
 \end{center}
\end{figure}

\begin{figure}[ht]
 \begin{center}
\mbox{\resizebox{!}{6cm}{\includegraphics{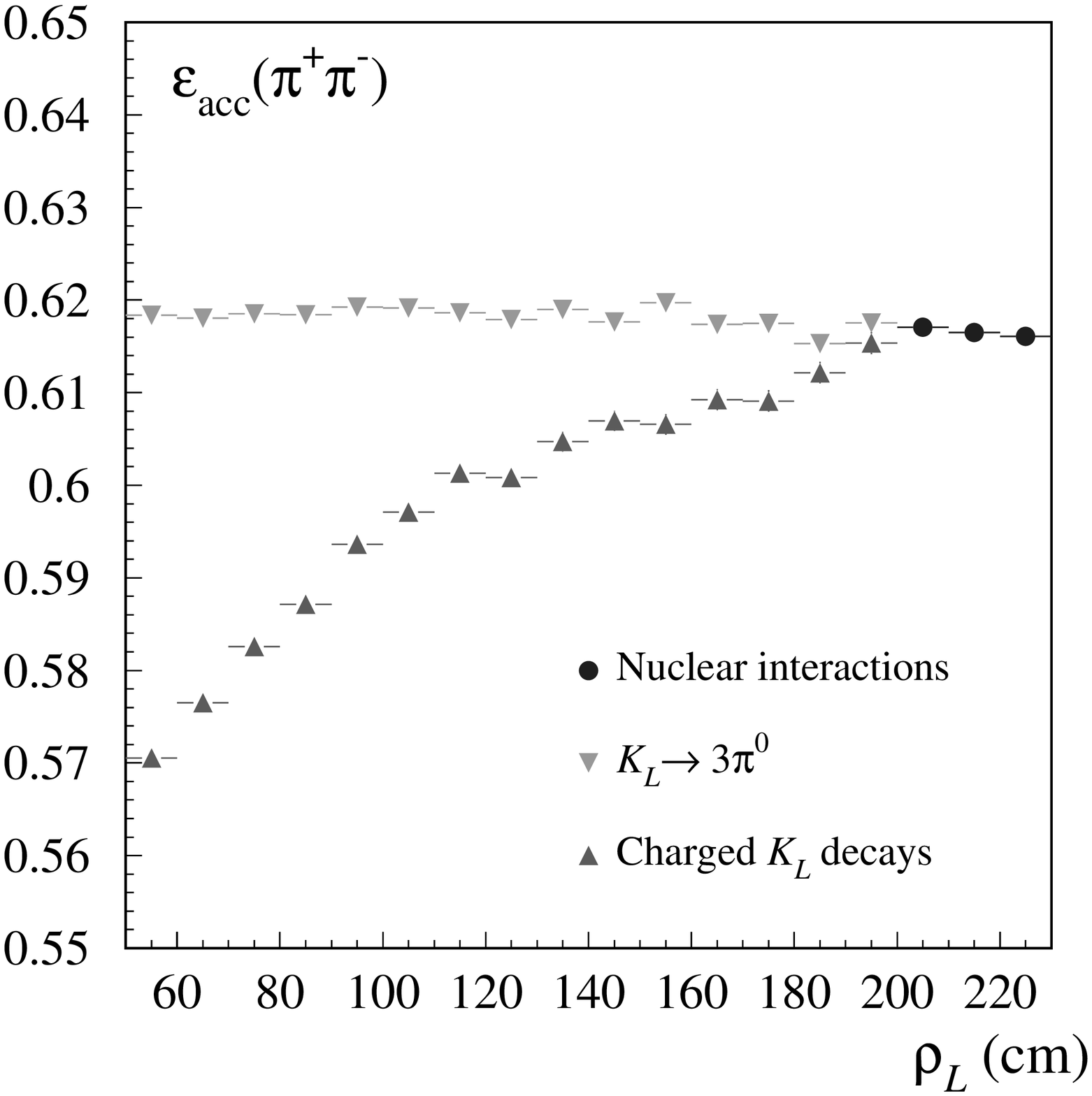}}}
  \caption{Acceptance of \DKSpippim\ as a function of \rT, for different \kl\ decay/interaction channels.}
    \label{fig:rtacce}
\end{center}
\end{figure}

Furthermore, \kcrmc\ and non-\kcrmc\ events have different topologies:
the first category is dominated by real \kl\ interactions in the EMC, with \bstar\
lying around the peak; the second category is mostly due to in-flight \kl\ decays before the EMC.
These two topologies also correspond to different \kl\ energy releases in the EMC, the latter topology being
much softer than the former (\Fig\ref{fig:evsbeta}).
If the \kcr\ tag is selected using a low value for the minimum energy cut 
($\ecr\!=\!125$\MeV), there is substantial contamination from in-flight \kl\ decays occurring
before the EMC. 
This is shown by the MC distribution of the transverse position (\rT) corresponding 
to the \kl\ decay or interaction producing the 
\kcr\ cluster (\Fig\ref{fig:rtspectra}). This contamination completely disappears when the cut 
is increased to 300\MeV.
Due to interference between \kl\ and \ks\ decay products, which undermines \ks\ reconstruction performance, 
the \ks\ signal acceptance is a function of the \kl\ decay mode and of the position \rT\ (\Fig\ref{fig:rtacce}).
Therefore, the signal
acceptance $a_{\kk}$ for the \kcrmc\ events, which are dominated by \kl\ interactions in the EMC,
is a few percent higher than that for non-\kcrmc\ events, $a_{\nk}$.

Finally, the selection efficiency of \Eq\ref{accesimple} is
expressed by combining the acceptances with the probabilities $\ek$ and $1-\ek$
for having a \kcrmc\ or non-\kcrmc\ event, respectively:
\begin{equation}
  \label{tnpn}
  \epsilon_{\rm sel} = \left[\ek\, a_{\kk}\, A +
  (1-\ek)\, a_{\nk}\, B \right]\,\epsilon_{\rm trg}\,\epsilon_{\rm CV}\,\epsilon_{\rm FILFO}.
\end{equation}
The fractions $A$ and $B$ are evaluated using data control samples, while
the efficiency $\ek$ is taken from MC (Sec.~\ref{tag}).
The acceptances $a_{\kk}$ and $a_{\nk}$ are evaluated using the MC, with data-driven corrections as explained
in Secs.~\ref{pippim} and \ref{piopio} for \pippim\ and \piopio\ events, respectively.
The efficiencies $\epsilon_{\rm trg}$, $\epsilon_{\rm CV}$, and $\epsilon_{\rm FILFO}$ 
are evaluated using data control samples 
as discussed in Sec.~\ref{trgeff}.
All of the efficiencies in \Eq\ref{tnpn} 
are to be understood as conditional probabilities, with each defined relative to the 
sample from the previous step in the analysis, according to the order in which they are applied.

The analysis is carried out for three different cuts on the
\kl\ cluster energy: $\ecr\!=\!125$, 200, and 300\MeV. 
The tagging efficiencies are very different in each case: $\ek\!\ab0.31$, 0.22, and 0.11,
respectively.
The fraction of \kl\ in-flight decays entering the selection varies significantly as well.
Moreover, some of the corrections applied and the related systematic uncertainties 
vary considerably with the cut value. This allows the robustness of the result to be tested.

The data were divided into 17 different samples following small changes in the machine energy.
The large number of events allowed a statistical error at the few per-mil level to be obtained 
for each single data period. Comparison of the independent measurements from each
data sample provides a stringent test of the validity of the corrections for possible
variations in the selection efficiency during data taking.
Results will be presented for each \kcr\ energy cut, averaging over all 17 samples.
The final result is obtained by choosing the value of \ecr\ which minimizes the total error.
Numerical details concerning all of the quantities involved in Eqs.~\ref{strategone} and~\ref{tnpn} 
are given in the following sections for a representative sample (no.\,10).
\subsection{Tagging efficiencies}
\label{tag}
This section concerns the evaluation of the quantities involved in the determination of the tagging 
efficiency: $A$, $B$, and \ek. 

The following parametrizations are used: $A\!=\!\sum_{n}\tn\pin$, $B\!=\!\sum_{n}\tn\pout$, where
\begin{itemize}
\item \tn\ is the \tzeroev\ spectrum, {\it i.e.}, the fraction of events in which \tzeroev\ is shifted by 
$n\times\tbunch$
with respect to the true collision time \tzeromc;
\item \pin\ is the probability that, given a found \kcrmc\ event, the \kcr\ tag is again found even after
the \tzeroev\ determination is shifted by $n\times\tbunch$; 
\item \pout\ is the probability that, in the absence of a found \kcrmc\ event, a \kcr\ tag is newly found after
the \tzeroev\ determination is shifted by $n\times\tbunch$. 
\end{itemize}
All of these quantities are taken from data control samples.

\begin{figure}[ht]
  \begin{center}
    \mbox{
     \resizebox{!}{6cm}{\includegraphics{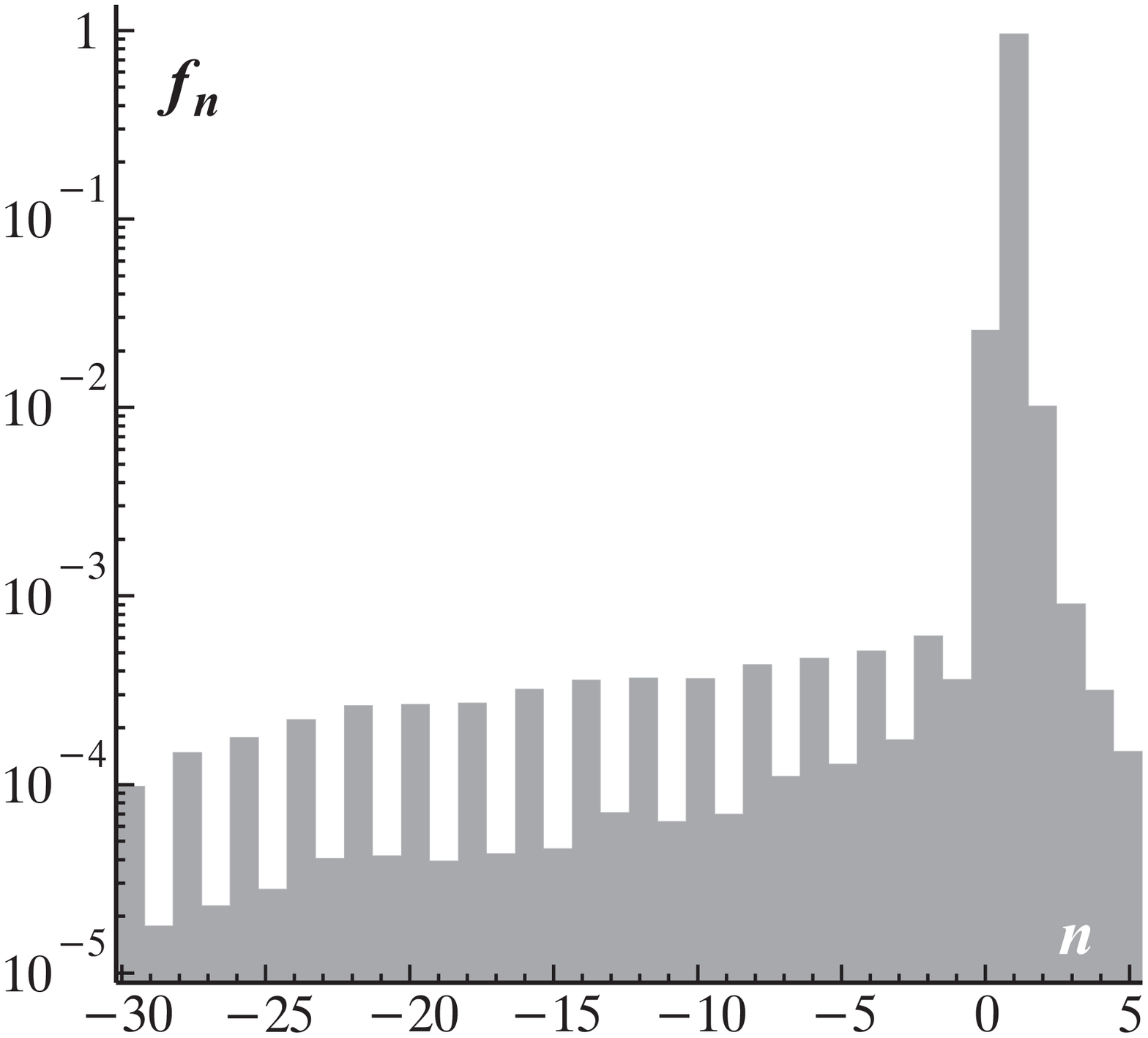}}
     }
    \caption{\tn\ spectrum for \KSpippim\ events.}
    \label{tnspectrum}
  \end{center}
\end{figure}

The \tzeroev\ spectrum (\tn) is evaluated for both \pippim\ and \piopio\ events after the signal selection 
requests have been applied.
For the charged mode, a subsample of \KSpippim\ events
is selected in which 
 both charged pions are associated to clusters. 
For each pion, an estimate of \tzeromc\ is obtained from the cluster time
and the time of flight calculated from the track parameters.
The robustness of this estimate is increased by requiring that 
both pions give the same result. The \tn\ spectrum is obtained as the normalized distribution of
 $(\tzeroev - \tzeromc)/\tbunch$ (\Fig\ref{tnspectrum}). 
As previously mentioned, \tzeroev\ overestimates the true collision time by one RF period 
\ab97\% of the
time. The negative tail of the spectrum shows peaks corresponding 
to the bunch-crossing times, and is dominated by events 
in which \tzeroev\ is determined
by a cluster from machine background
occurring at random with respect to the collision time.
For \piopio\ events the situation is much simpler, because the request of having at least three prompt clusters 
is fulfilled only if 
$\tzeroev\!=\!\tzeromc$. Therefore $\tn$ is negligible for $n\neq0$, and the values 
 $A(\piopio)\!=\!1$ and $B(\piopio)\!=\!0$ are used.

\begin{figure}[ht]
 \begin{center}
  \begin{tabular}{ccc}
    \resizebox{!}{3.7cm}{\includegraphics{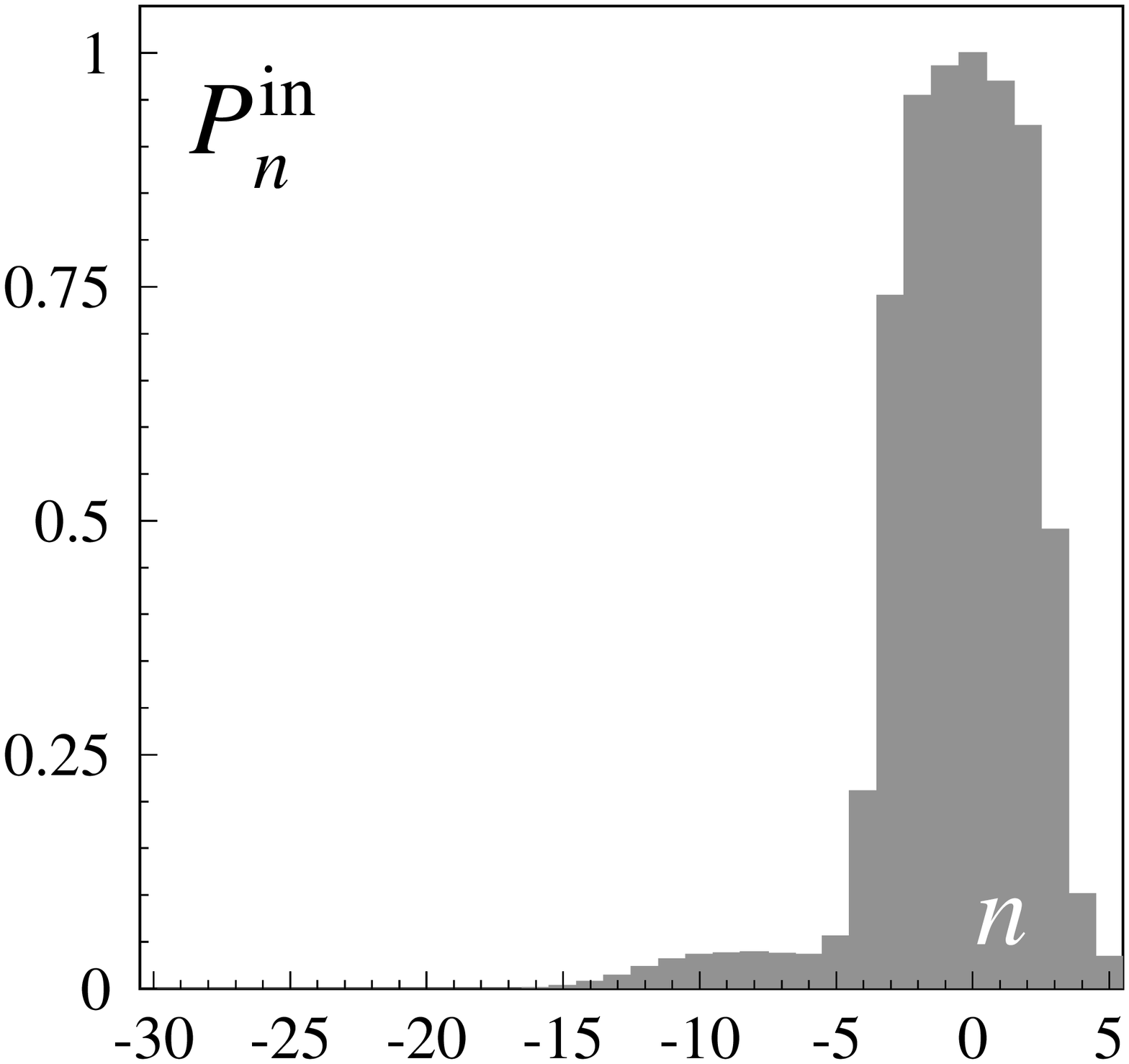}} & \,\,\,\,\, &
    \resizebox{!}{3.7cm}{\includegraphics{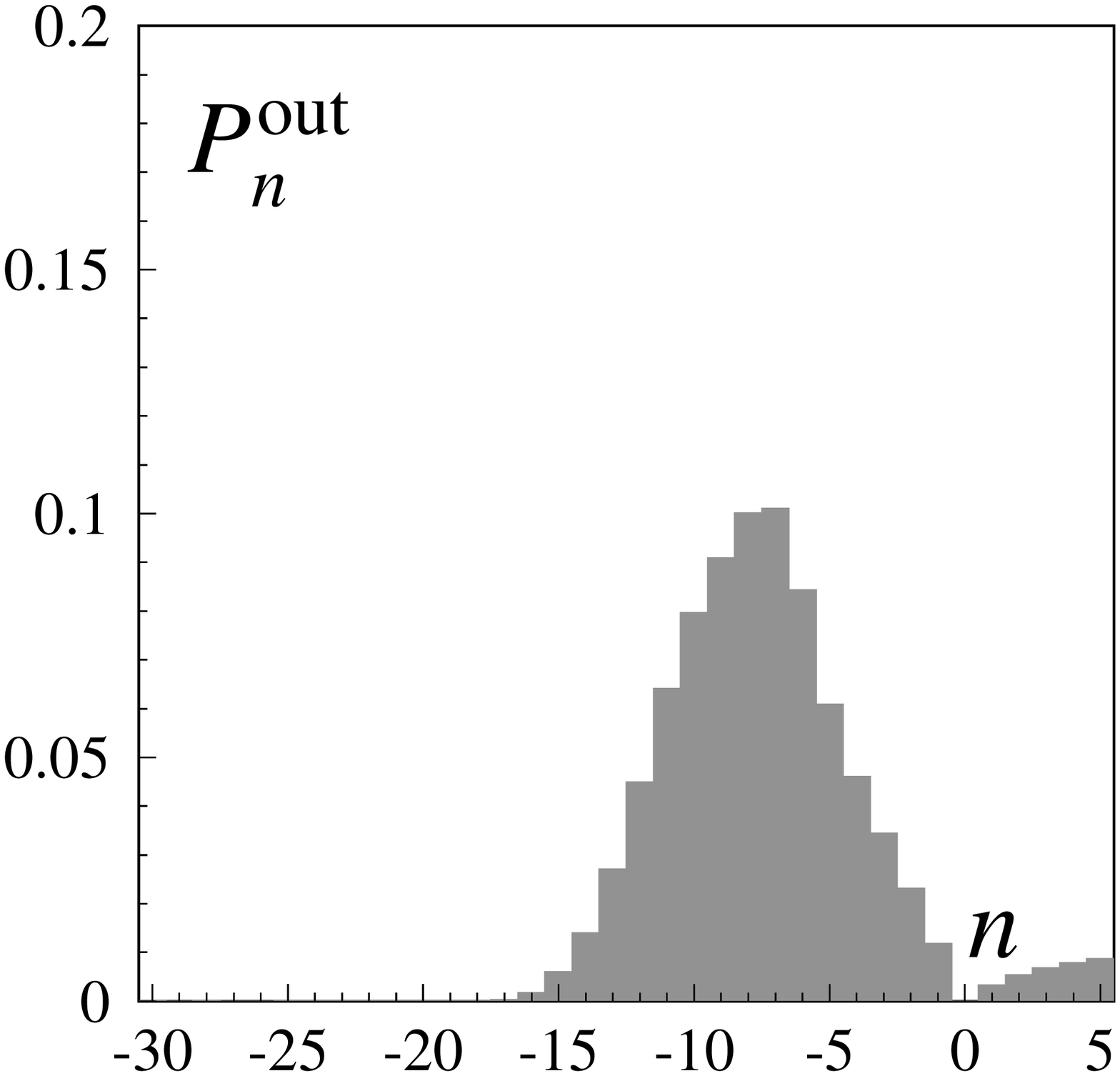}}
  \end{tabular}
  \caption{Probabilities \pin\ (left) and \pout\ (right) as a function of the \tzeroev\
    shift ($n$). By definition, $P^{\rm \,in}_{0}\!=\!1$ and $P^{\rm \,out}_{0}\!=\!0$.}
    \label{fig:pinpout}
\end{center}
\end{figure}

For the charged mode, the probabilities \pin\ and \pout\ are needed for the evaluation of $A$ and $B$.
For this purpose, a sample of events selected on the basis of a reconstructed \KSpippim\ decay 
(without reference to the \kcr\ tag) is used.
The estimate of the true collision time \tzeromc\ described above is used
to divide these events into \kcrmc\ and non-\kcrmc. The \tzeroev\ value is then artificially shifted
by $n\times\tbunch$ with respect to \tzeromc. For \kcrmc\ events,
the probability \pin\ of still finding the \kcr\ is evaluated, as is the probability \pout\ of finding
a \kcr\ not originally present for non-\kcrmc\ events. 
The probabilities \pin\ and \pout\ are shown as a function of the \tzeroev\ shift in 
\Fig\ref{fig:pinpout}.
From the probabilities \tn, \pin, and \pout\ the fractions
$A$ and $B$ are calculated.
The results for \pippim\ events are listed in
\Tab\ref{tagtab}. The value of $A(\pippim)$ 
increases with \ecr, reaching
\ab99\% for $\ecr\!=\!300$\MeV. 
The tails of the \bstar\ spectrum are indeed suppressed 
by increasing the \kcr\ energy cut, as shown in \Fig\ref{fig:evsbeta}. 
This reduces the acceptance losses due to incorrect \tzero\ determination. 
The maximal variation of $A(\pippim)$ during data taking is \ab1\%.

Possible biases in the estimate of the \tn\ spectra have been checked using the
 MC, by evaluating the ratio $A^{\rm true}/A$.
Here, $A^{\rm true}$ is evaluated from MC truth
and $A$ is evaluated with the same method used for data. 
The above ratio is applied as a correction to the estimate of $A$ for data; the systematic 
error, taken
as 100\% of the correction, 
amounts to $0.25\times 10^{-3}$ for $\ecr\!=\!300$\MeV. Given the small value of 
$B(\pippim)$, no correction is applied.
A similar comparison with MC truth allows a systematic error of 
$\ab0.4\!\times\!10^{-3}$ to be assigned to the assumption $A(\piopio)=1$. 
The total systematic error
from the evaluation of the \tn\ spectra and of the probabilities \pin, \pout\ is $0.45\!\times\!10^{-3}$ at 
$\ecr\!=\!300$\MeV\ (see \Tab\ref{systtab}).

\begin{figure}[ht]
 \begin{center}
  \begin{tabular}{ccc}
    \resizebox{!}{3.7cm}{\includegraphics{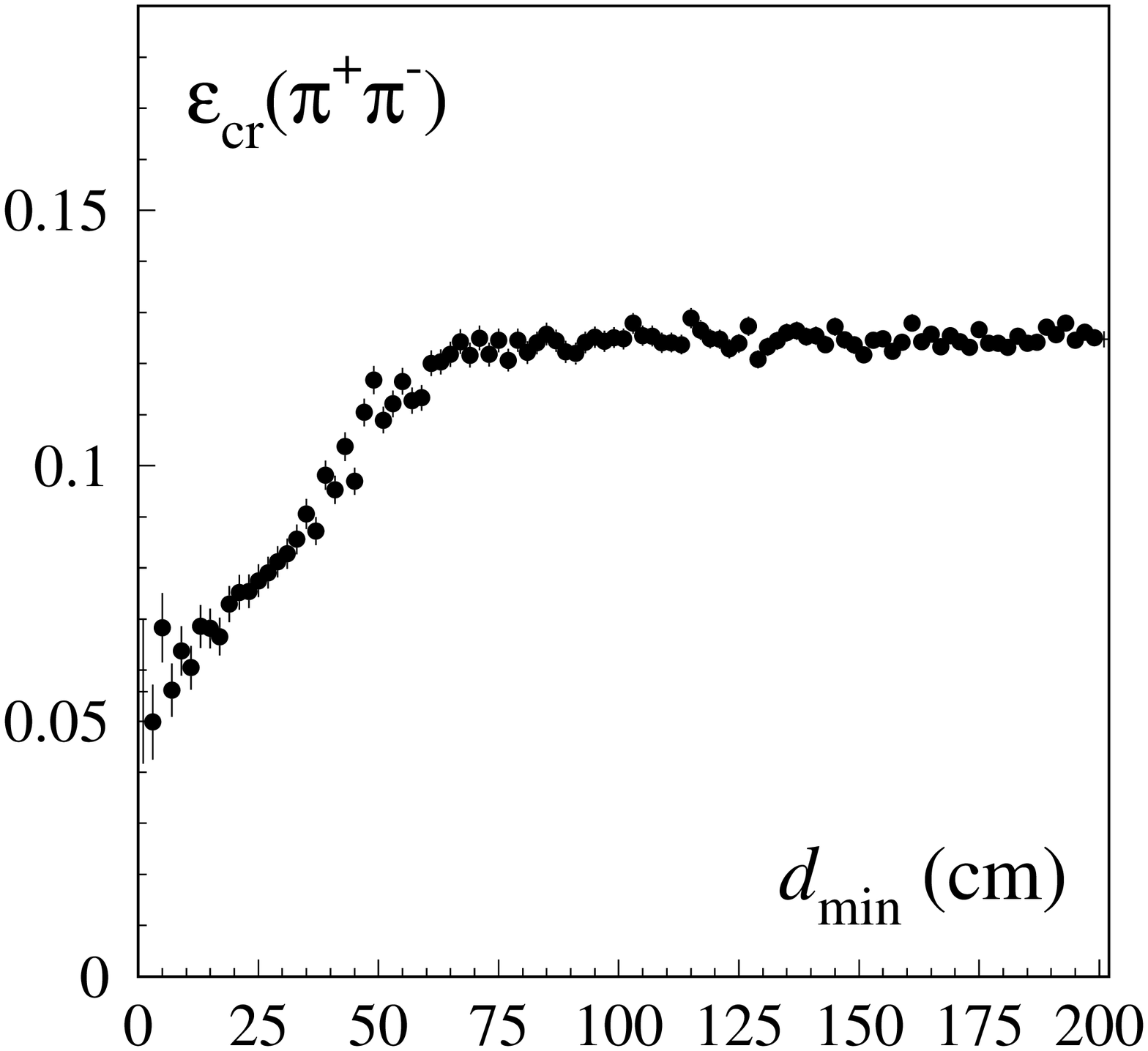}} & \,\,\,\,\, &
    \resizebox{!}{3.7cm}{\includegraphics{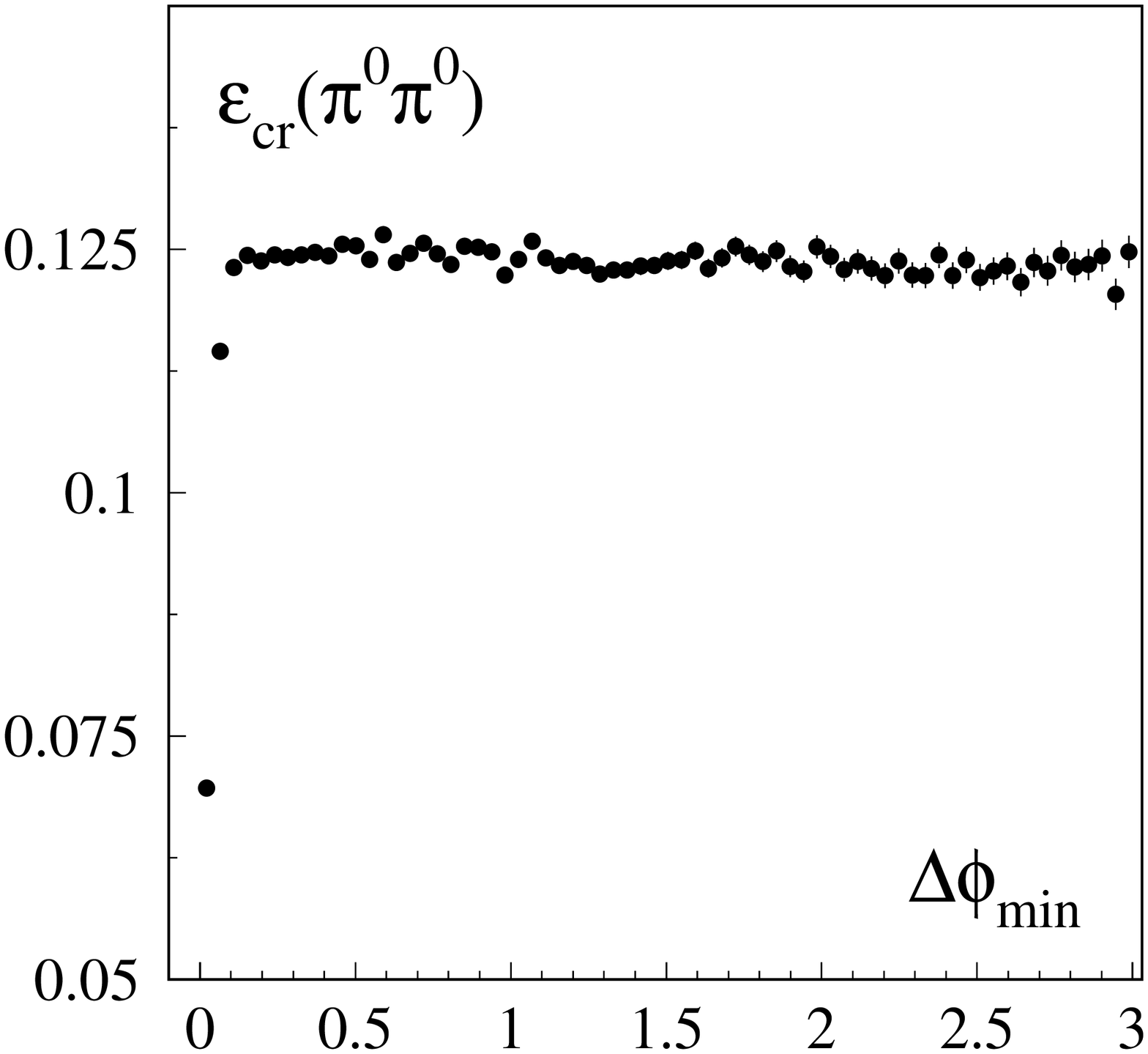}}
  \end{tabular}
  \caption{$\ek(\pippim)$ as a function of \dmin\ (left) and $\ek(\piopio)$ as a function
of $\Delta\phi_{\rm min}$ (right), for $\ecr\!=\!300$\MeV. The efficiencies shown have been
obtained using control samples of signal 
events in which at least one \ks\ decay product reaches the barrel.}
    \label{fig:efficr}
\end{center}
\end{figure}

\begin{figure}[ht]
 \begin{center}
  \begin{tabular}{ccc}
    \resizebox{!}{3.7cm}{\includegraphics{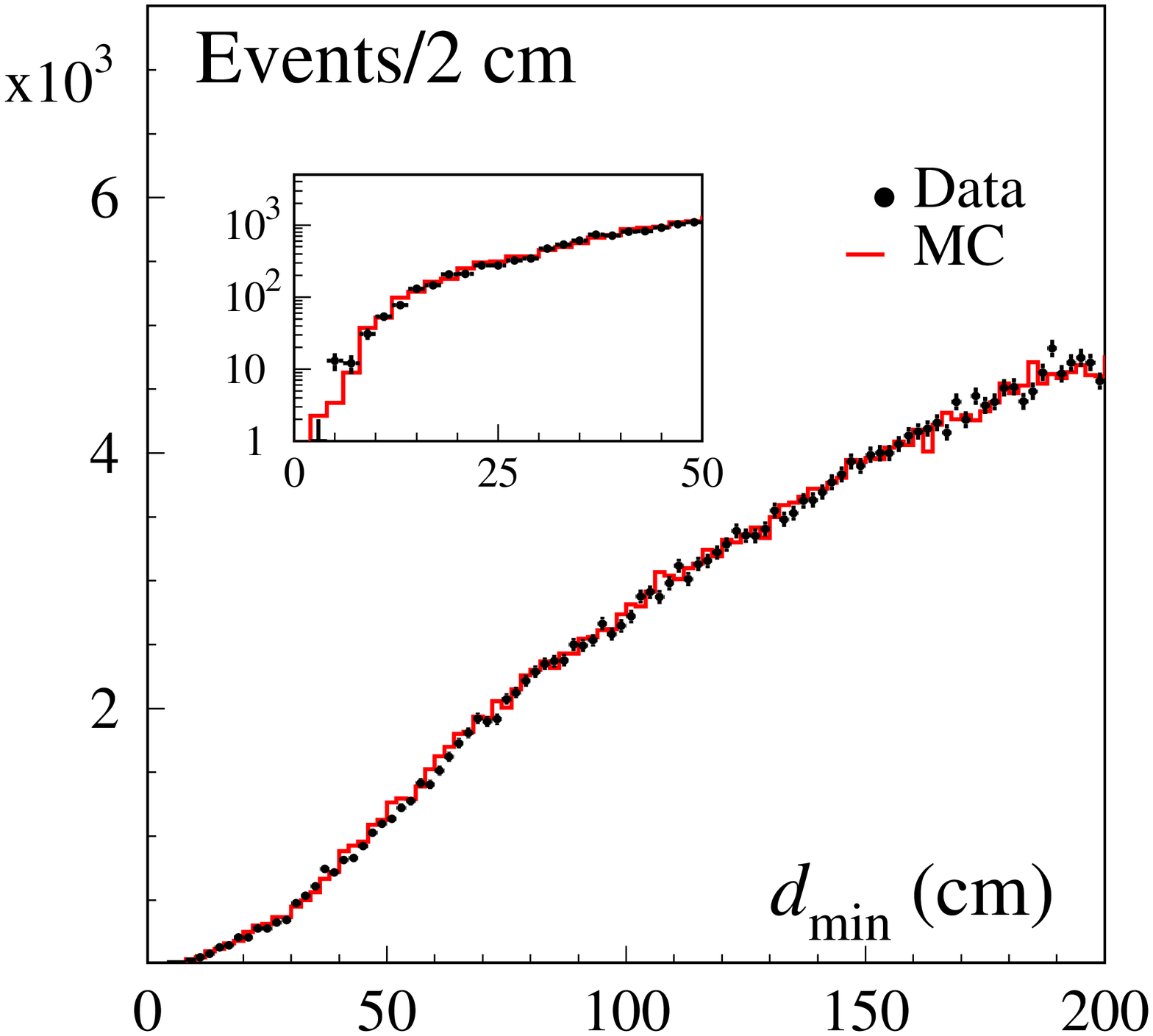}} & \,\,\,\,\, &
    \resizebox{!}{3.7cm}{\includegraphics{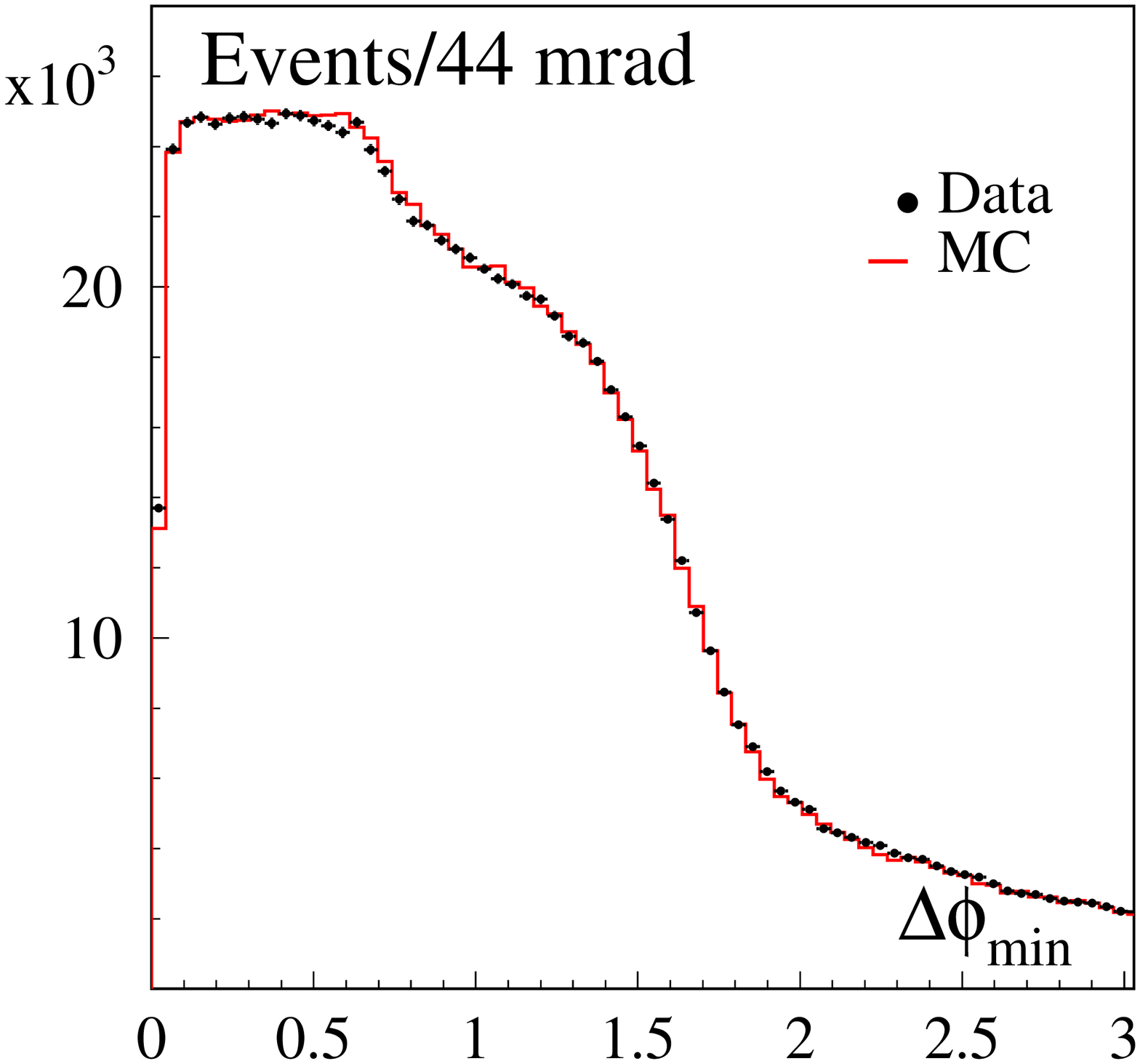}}
  \end{tabular}
  \caption{Comparison between data and MC distributions of \dmin\ for \pippim\ events (left) and $\Delta\phi_{\rm min}$
for \piopio\ events (right).}
    \label{fig:sovrocr}
\end{center}
\end{figure}

When computing the ratio between \pippim\ and \piopio\ selection efficiencies 
(see Eqs.~\ref{strategone} and \ref{tnpn}) the values of
the \kcrmc\ efficiencies $\ek(\pippim)$ and $\ek(\piopio)$ are needed; since
$B(\pippim)\!\ab0.3\%$ and $B(\piopio)\!=\!0$, the ratio of selection efficiencies
depends on the ratio $\ek(\pippim)/\ek(\piopio)$ rather than on the \ek\ values for each channel.
This ratio varies with \ecr, 
ranging from \ab1.003 at $125$\MeV\ up to \ab1.014 at $300$\MeV\ 
(\Tab\ref{tagtab}). 
This is due to the geometrical overlap in the EMC between \ks\ daughter particles and the \kl, 
which affects the \kcr\ reconstruction 
efficiency in a manner dependent on the decay channel.
For \pippim\ events, the \kcrmc\ efficiency drops when the pions get closer to the \kl\ 
because of the higher probability of associating 
the \kl\ cluster to one pion track; for \piopio\ events, a drop is observed  
when a \ks\ photon and the \kl\ hit the same calorimeter 
cell, thus spoiling the cluster reconstruction. 
These effects have been studied using MC control samples of signal events in which
at least one \ks\ decay product reaches the EMC barrel. The effects are then visible in 
the dependence of $\ek(\pippim)$ on the minimum distance \dmin\ 
between the \kl\ and the closest \ks\ decay product on the barrel and in the
dependence of $\ek(\piopio)$ on the minimum angular distance $\Delta\phi_{\rm min}$ in the transverse plane
(\Fig\ref{fig:efficr}). 
Biases are present only when
\ks\ daughter particles enter the EMC close to the \kl\ impact point.
The reliability of the MC in reproducing this overlap effect is checked by comparing data and MC 
distributions of \dmin\ and $\Delta\phi_{\rm min}$ for events with a \kcr\ tag found (\Fig\ref{fig:sovrocr}).
The ratio of data and MC distributions is constant in the region safe from overlap effects. A significant discrepancy
 is only present for \pippim\ events when $\dmin\!<\!10$\cm. 
The MC evaluation of $\ek(\pippim)/\ek(\piopio)$ is corrected by scaling the number of \kcr\ 
events found for small \dmin\ values according to the ratio measured for data.
The systematic error, taken as 100\% of the correction, amounts to $\ab0.44\!\times\!10^{-3}$ 
for $\ecr\!=\!300$\MeV\ (\Tab\ref{systtab}). 

\begin{table}[ht!]
  \begin{center}
\renewcommand{\arraystretch}{1.2}    
    \begin{tabular}[c]{|c|c|c|c|c|}\hline
 \multicolumn{2}{|c|}{\ecr\ value} &  125\MeV  & 200\MeV &  300\MeV     \\ \hline 
 \pippim\ &   $A$                    &   0.9634(1) & 0.9866(1) & 0.9933(1)  \\ 
          &   $B (\times 10^{-3})$   &   3.4489(6) & 1.6675(1) & 0.71563(3) \\ 
          &   \ek\                   &   0.3106(2) & 0.2231(2) & 0.1082(2)  \\ \hline 
 \piopio\ &   \ek\                   &   0.3097(3) & 0.2217(3) & 0.1067(2)  \\ \hline 
    \end{tabular}
  \end{center}
  \caption{Tagging probabilities entering into the evaluation of the selection efficiency (\Eq\ref{tnpn})
 for \pippim\ and \piopio\ events, for data sample no.\,10 and 
 minimum \kcr\ energies of 125, 200, and 300\MeV. Statistical errors on the last 
digit are
shown in parentheses.}
  \label{tagtab}
\end{table}

\subsection{Acceptance and purity for \DKSpippim}
\label{pippim}
\begin{figure}[ht]
\begin{center}
    \mbox{
     \resizebox{!}{8cm}{\includegraphics{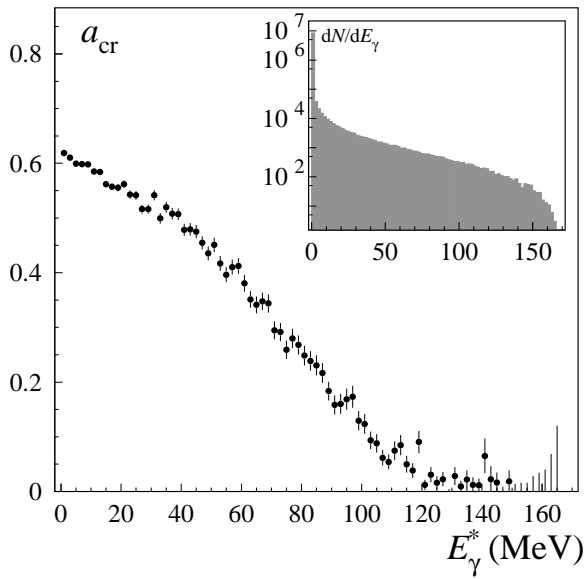}}
     }
\caption{Acceptance $a_{\kk}$ for \DKSpippim\ as a function of the center-of-mass photon energy.
The photon spectrum used in the simulation is shown in the inset.}
\label{fig:spettrophoton}
 \end{center}
\end{figure}

The \pippim\ acceptance is evaluated from MC.
Since no cut is applied on the $\pi\pi$ invariant mass, the selection
includes $\ks\!\to\!\pippim\gamma$ events with
photon energies up to the end point
(\ab160\MeV\ in the \ks\ rest frame). 
However, due to the fact that both pion tracks must extrapolate to the calorimeter without spiralling, 
the acceptance depends on 
the photon energy: the harder the photon in the final state, the higher the probability that one of the
pion tracks spirals in the chamber
 before reaching the EMC. 
The MC simulation includes final-state radiation~\cite{radiativearticle}.
The acceptance obtained by MC is shown in \Fig\ref{fig:spettrophoton} as a function of the
photon energy $E^{\ast}_{\gamma}$ in the \ks\ rest frame.
The simulated spectrum is also shown in the inset of \Fig\ref{fig:spettrophoton}. The fraction of events
with $E^{\ast}_{\gamma}\!>\!20$ and 50\MeV\ are $7.0\!\times\!10^{-3}$ 
and $2.5\!\times\!10^{-3}$, respectively, in
excellent agreement with the measured values $(7.10\pm0.22)\!\times\!10^{-3}$ 
and $(2.56\pm0.09)\!\times\!10^{-3}$~\cite{ramberg:93}.
The MC calculation thus provides a fully inclusive   
acceptance, which is \ab0.3\% lower than that obtained with a pure $\ks\to\pippim$ simulation. 

A  crucial issue when evaluating the \pippim\ acceptance is to correctly reproduce the 
DC tracking efficiency, including all possible variations correlated with the level of machine
background and with the hardware performance of the apparatus.
For this purpose, accidental background hits in the DC are extracted from real
$\epm\!\to\!\gamma\gamma$ events and are overlaid with the simulated events; moreover, the
measured hardware hit efficiency is used to sample the MC hit generation~\cite{OFFnim}.
To take into account residual
differences in the tracking efficiencies for data and MC, 
the acceptance calculation
is performed by weighting the contribution of each single pion 
with the ratio
$\etrk^{\rm data}/\etrk^{\rm MC}$. 
The single-track efficiencies \etrk\ for data and MC are evaluated from a subsample of 
\KSpippim\ events tagged by a \kcr. Using the \ks\ momentum $\vecbf{p}_{\ks}$ as determined
from $\vecbf{p}_{\phi}$ and by the \kl\ flight direction, it is indeed possible
to identify the \pippim\ final state by selecting a single pion track (``tagging'' track) with the
expected momentum in the \ks\ rest frame: $201\!\leq\!p^{\ast}_{\rm tag}\!\leq\!209$\MeVsuc.
This selection reduces background to a negligible level, while at the same time providing a good 
estimate of the momentum of the other pion: $\vecbf{p}_{\rm other}\!=\!\vecbf{p}_{\ks}-\vecbf{p}_{\rm tag}$. 
The single-track efficiency is then obtained by counting the fraction of times in which a second pion track 
is found; it is evaluated in bins of transverse and longitudinal momenta, separately for each particle
charge. This method takes into account not only differences in \etrk\ for data and MC, but also 
differences between the real and simulated
nuclear interaction cross sections for the pions.

The MC calculation is also corrected for data-MC differences in 
the efficiency \eto\ for a single pion with impact on the calorimeter to provide a \tzero\ cluster.
This is evaluated using various data control samples (\KSpippim, \Dphipippimpio) as a function of the track momentum and
the angle of incidence on the EMC, distinguishing between $\pi^{+}$ and $\pi^{-}$ tracks (or $\mu^{+}$ and $\mu^{-}$
tracks, in case of in-flight pion decays), and separately treating tracks reaching the 
barrel or the endcaps.

The values of  $a_{\kk}$ and $a_{\nk}$ are listed in \Tab\ref{pippimacce}, 
together with the number of events selected as \pippim. The errors quoted 
on $a_{\kk}$ and $a_{\nk}$ 
are due to the statistics of the MC sample and of 
the control samples used for the efficiency determination. The maximal variation of the acceptance 
during data taking is \ab2\%, and is due to variations in the machine
operating conditions (background levels and center of mass energy).   
The acceptance $a_{\nk}$ is \ab3\% lower than $a_{\kk}$
for all values of \ecr, because of
the presence of \kl's decaying into charged particles before reaching the EMC, which disturb
the reconstruction of \ks\ pion tracks as discussed in section~\ref{effischeme}.
The value of $a_{\kk}$ increases by $0.8\!\times\!10^{-3}$ as the \ecr\ cut is moved from 125 to 300\MeV.
This is due to the contamination from late \kl\ decays
in the \kcrmc\ sample (\Fig\ref{fig:rtspectra}), which is suppressed when the \kcr\ energy cut is raised. 
The above variation is taken as a conservative estimate
of the systematic error from the simulation of this \ks-\kl\ interference.
A further contribution to the systematic 
error comes from the \tzero\ efficiency correction, \eto; it is estimated 
by MC as the difference between the result of the method described above
and the MC truth. A non-zero difference is found and 
is ascribed to interference between the two decay products of the \ks, 
which is not correctly taken into account by the above method.
The difference is $1.4\!\times\! 10^{-3}$ at $\ecr\!=\!300$\MeV\ (\Tab\ref{systtab}). This value
is both applied as a correction and taken as a conservative estimate of the systematic error.

The purity $C$ of the \pippim\ sample is estimated by MC to be \ab0.9989 and is independent
of \ecr\ (see \Tab\ref{pippimacce}). Two sources contribute to the background contamination: 
\ks\ decays to semileptonic final states 
($\ab0.7\!\times\!10^{-3}$) and $\Dphipippimpio$ decays ($\ab0.4\!\times\!10^{-3}$). 
Semileptonic decays  are able to satisfy with high efficiency the loose kinematic criteria
used to select \pippim\ events.
Events with $\Dphipippimpio$ decays enter the selection when an early accidental
cluster establishes \tzero\ and one of the two high-energy photons from the $\pi^{0}$ is 
erroneously selected as the \kcr. The systematic error on the purity comes from the uncertainty on the BR's for 
the decays
involved and, for the $\Dphipippimpio$ contribution, from the uncertainty on the rate \Racc. 
The error from these sources is 
$0.1\!\times\!10^{-3}$ at $\ecr\!=\!300$\MeV\ (\Tab\ref{systtab}).

\begin{table}[ht!]
  \begin{center}
\renewcommand{\arraystretch}{1.2}    
    \begin{tabular}[c]{|c|c|c|c|}\hline
 \ecr\ value            &  125\MeV    & 200\MeV &  300\MeV          \\ \hline 
 $N$                  &  1,218,000  &  907,400  &  490,900           \\ \hline
 $a_{\kk}$            &   0.6187(5) & 0.6192(6) & 0.6195(8)           \\ 
 $a_{\nk}$            &   0.5968(5) & 0.5991(5) & 0.6016(5)           \\ 
 $C$                  & 0.99882(4) & 0.99891(4) & 0.99886(6)          \\ \hline
    \end{tabular}
  \end{center}
  \caption{Values for the observed yield, the acceptance, and the purity of the \pippim\ selection, 
for data sample no.\,10 and minimum \kcr\ energies of 125, 200, and 300\MeV. Statistical errors  on the last 
digit are
shown in parentheses.}
  \label{pippimacce}
\end{table}

\subsection{Acceptance and purity for \DKSpiopio}
\label{piopio}
The \DKSpiopio\ acceptance is evaluated from MC. 
To take into account
data-MC differences in the cluster efficiency \ecl\ for low-energy photons,
the acceptance calculation is performed by weighting 
each photon with the ratio \ecl(data)/\ecl(MC).
The single-photon detection efficiencies are evaluated 
from control samples of \Dphipippimpio\ events, which are selected
using DC information only: two
tracks with opposite charge from the interaction point (IP) are required, with
a missing four-momentum $p_{\Ppio}\!=\!p_{\phi} - p_{\pi^+} - p_{\pi^-}$  
compatible with the \Ppio\ mass hypothesis.
A photon from \Ppio\ decay is identified (``tagging'' photon, $\gamma_1$) as a
cluster with  time of flight and energy 
in an appropriate interval around the expected values. 
The energy is derived from the \Ppio\ momentum 
and the position of the cluster for $\gamma_1$ using the relation
$E_{\gamma 1}\!=\!m^2_{\Ppio}/2(E_{\Ppio}-p_{\Ppio}\cos{\theta_{\Ppio \gamma 1}})$.
The above selection provides a good estimate of 
the momentum of the second photon, $\vecbf{p}_{\gamma 2}\!=\!\vecbf{p}_{\Ppio}-\vecbf{p}_{\gamma 1}$. 
The photon efficiency \ecl\ 
is then obtained by counting the fraction of times in which the second photon is found in a 
cone around the expected direction. The result is evaluated in bins in the expected polar angle and
energy; photons from \Dphipippimpio\ events have a wider energy 
spectrum than that for \DKSpiopio\ events, so that the efficiency can be successfully 
evaluated up to the end point, $E_{\gamma}\ab300\MeV$.

The values of $a_{\kk}$ are listed in \Tab\ref{piopioacce},
together with the number of events selected as \piopio. 
The maximal variation in the acceptance during data taking is \ab1\% and is due to
variations in the machine background.
The various sources of systematic uncertainty on the acceptance evaluation are discussed below.

\begin{figure}[ht]
\begin{center}
    \mbox{
     \resizebox{!}{4cm}{\includegraphics{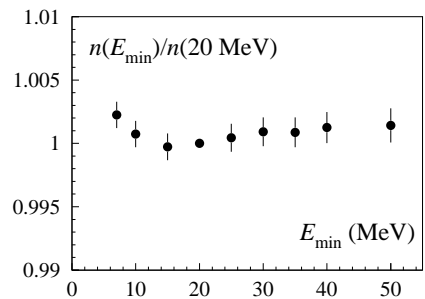}}
     }
\caption{Variation of the number of \DKSpiopio\ events relative to that for a 20-\MeV\ cut,
as a function of the minimum energy cut $E_{\rm min}$.
Each number is obtained correcting the event count with the corresponding efficiency.}
\label{fig:stabilitypopo}
 \end{center}
\end{figure}

A first contribution to the systematic uncertainty on the photon counting arises 
from uncertainty in the data-MC cluster-efficiency correction. This has been evaluated by 
varying the cut on the minimum cluster energy, $E_{\rm min}$, 
from the default value of 20\MeV\ to values 
between 7 and 50\MeV\ and
checking the stability of the number of selected events after efficiency corrections, 
$n(E_{\rm min})$. 
When the cut is moved from 7 to 50\MeV, the acceptance decreases by \ab18\%. The data-MC cluster
efficiency correction is \ab0.9965 with the cut at 7\MeV, and is negligible with the cut at 50\MeV.
The variation of $n(E_{\rm min})$ normalized to $n(20\MeV)$
is shown as a function of $E_{\rm min}$ in \Fig\ref{fig:stabilitypopo}.
The associated fractional systematic error is $0.5\!\times\!10^{-3}$.

An additional systematic uncertainty in photon counting arises from a data-MC difference in the 
probability for a photon to produce more than one prompt cluster (``splitting'').
If this occurs, an event with only two real prompt photons might
be accepted as a three-prompt-photon event. The relative bias induced in the acceptance is proportional to the
difference between data and MC splitting probabilities: 
\begin{equation}
\label{split}
 \frac{\Delta a_{\kk} }{a_{\kk}} 
= \left( \pspli^{\rm data} - \pspli^{\rm MC } \right) \times 
\frac{2{N_2}}{{N_{\ge 3}}},
\end{equation}
where $N_i$ ($N_{\ge i}$) represents the number of \DKSpiopio\ events with $i$ ($\ge i$) prompt photons. 
The splitting probabilities are evaluated for data and MC using events with a \kcr\ and five
prompt clusters. In this sample, there is always either one split or one accidental cluster. 
The splitting probability is 
then evaluated as $\pspli\!=\!N_{\rm split}/(4N_4)$, where
$N_{\rm split}$ is the number of five-prompt events in which
a pair of clusters closer than 80\cm\ is found. 
The results for data and MC are $\pspli^{\rm data}\!\ab2.7\!\times\!10^{-3}$
and $\pspli^{\rm MC}\!\ab1.4\!\times\!10^{-3}$. Given a ratio $N_2/N_{\ge 3}\!\ab0.09$,
the bias on the acceptance from \Eq\ref{split} is
$0.22\!\times\!10^{-3}$; this is taken both as a correction and as an estimate of the systematic
uncertainty due to this effect.

Photons from \DKSpiopio\ have a probability of $\ab4\!\times\!10^{-3}$ to convert to an \epm\ pair
before entering the DC volume. Moreover, there is a probability of $\ab2.4\%$ that at least
one \Ppio\ undergoes a Dalitz decay~\cite{PDBook}.
These two categories of events produce at most three prompt clusters and are therefore
selected with a lower efficiency $a_{\kk}^{\rm pair}\!\ab0.67$ instead of $a_{\kk}\!\ab0.89$. The \piopio\
acceptance, which is averaged over the populations with and without \epm\ pairs in the final state is 
therefore subject to error if the MC does not reproduce the real $\gamma$-conversion cross section
(the uncertainty due to the BR for the Dalitz decay is negligible).
This effect has been checked by searching for  
for tracks from the IP in events selected as \piopio\ in data and MC. 
If an $e^+e^-$ pair is produced, at least one track is reconstructed with a probability 
$p_{\rm trk}\!\ab0.74$. 
Having measured for data and MC the fraction $f_{\rm trk}$ of events with at least one track pair 
from the IP, the correction to the acceptance is evaluated as follows:
\begin{equation}
\label{couple}
\frac{\Delta a_{\kk} }{a_{\kk}} = 
\frac{f_{\rm trk}^{\rm data} - f_{\rm trk}^{\rm MC}}{p_{\rm trk} a_{\kk}^{\rm pair}}  
\times
\left(a_{\kk}-a_{\kk}^{\rm pair}\right).
\end{equation}
The difference $f_{\rm trk}^{\rm data} - f_{\rm trk}^{\rm MC}$ is $\ab10^{-3}$. 
This results in a $0.38\!\times\!10^{-3}$ bias on the acceptance, which is taken both as a correction and as an 
estimate of the systematic uncertainty due to photon conversion.

The total systematic error on the acceptance due to ``cluster counting'' effects is therefore 
$0.66\!\times\!10^{-3}$ at $\ecr\!=\!300$\MeV\ (\Tab\ref{systtab}).

In addition to the above effects,
the consequences of possibly incorrect \tzero\ estimates have been considered. An error on the \tzero\
results in an incorrect evaluation 
of the time of flight for each photon and causes the \piopio\ event to be lost. 
This can occur due to the presence of machine background clusters,
which determine the value of \tzeroev\ in 1-2\% of the events.
The uncertainty in \Racc\ (Sec.~\ref{setup}) therefore gives rise to a systematic error on the acceptance 
for \piopio\ events.
However, the acceptance for \pippim\ events is also affected by an error on \Racc, because 
drift times are wrongly evaluated when \tzero\ is incorrect.
The two effects partially cancel out when evaluating the ratio of \pippim\ and \piopio\ acceptances, leaving
a residual systematic error of $0.52\!\times\!10^{-3}$ at $\ecr\!=\!300$\MeV\ (\Tab\ref{systtab}).

When the \tzeroev\ determination is incorrect because two photons
hit the same calorimeter cell, or because
one photon cluster overlaps with a noisy EMC channel, a further loss of \piopio\ events occurs.
 In such cases,
the time of the \tzero\ cluster is badly reconstructed.
The fraction of events lost because of these mechanisms is \ab1\%. The associated correction has been 
evaluated from data samples of
\DKSpiopio\ events tagged by \DKLpippimpio\ decays in the DC, which can be selected independently of the 
\tzeroev\ determination.
The corresponding systematic error is $0.61\!\times\!10^{-3}$ (\Tab\ref{systtab}).

The \piopio\ sample is contaminated mainly by $K^+K^-$ events in which one of the two kaons
undergoes a decay to $\pi^{\pm}\piopio$ near the origin, while the other decays to $\Ppio$'s within the
DC. If the flight path of this second kaon is between \ab90 and \ab160\cm, 
one of the two photons from a $\Ppio$ decay can be 
 taken as a \kcr. This probability for this to occur strongly decreases with \ecr.
The purity $C$ is evaluated from MC and depends on \ecr\ as shown in \Tab\ref{piopioacce}.
A systematic error on this estimate comes from the uncertainties on the BR's 
involved in the decay chains and from the acceptance for
 $K^{\pm} \to \pi^{\pm}\piopio$. The uncertainty is $0.35\!\times\!10^{-3}$ at $\ecr\!=\!125$\MeV\ 
and negligible at $\ecr\!=\!300$\MeV\ (\Tab\ref{systtab}).
A minor source of background, also included in $C$, is due to events in which multiple clusters from
 machine background generate both 
 the \kcr\ tag and 
 three prompt clusters. The residual contamination is evaluated using data;
it is $0.13\!\times\!10^{-3}$ at $\ecr\!=\!125$\MeV\ and decreases by a factor of two at 
$\ecr\!=\!300$\MeV.
The systematic error due to these events is conservatively estimated to be equal to the contamination 
itself.
\begin{table}[ht!]
  \begin{center}
\renewcommand{\arraystretch}{1.2}    
\begin{tabular}[c]{|c|c|c|c|}\hline
  \ecr\ value            &  125\MeV & 200\MeV &  300\MeV  \\ \hline 
 $N$     & 811,800     &  587,700   &  312,900     \\ \hline
 $a_{\kk}$             & 0.8905(7)  & 0.8911(8) & 0.8910(9)   \\ 
 $C$                   & 0.9940(1) & 0.99761(8) & 0.99938(6)  \\   \hline
    \end{tabular}
  \end{center}
  \caption{Values for the observed yield, the acceptance, and the purity of the \piopio\ selection,
for data sample no.\,10 and minimum \kcr\ energies of 125, 200, and 300\MeV. Statistical errors on the last 
digit are 
shown in parentheses.}
  \label{piopioacce}
\end{table}
\subsection{Trigger, cosmic-ray veto, and offline filter efficiencies}
\label{trgeff}
The trigger efficiency for each channel is obtained from data. 
The trigger requires at least two fired sectors in the EMC and this condition can be satisfied
by \ks\ decay products or by the \kcr\ alone. The idea is therefore
to extract the probability $P_{L(S)}^{(i)}$ for the \kl(\ks) to fire $i$ trigger 
sectors by requiring that
the trigger condition be satisfied by the set of \ks(\kl) clusters, which are identified on the basis of the
time of flight. 
The trigger efficiency is then calculated by combining
\ks\ and \kl\ trigger sector probabilities. \kl\ interactions always fire at least one sector, so 
$P_L^{(2)}\!=\!1-P_L^{(1)}$. Events  
are lost when only one sector is fired by the \kcr\ ($P_L^{(1)}\!\ab60\%$) 
and no \ks\ decay product complements the \kcr\ cluster to satisfy the trigger:
\begin{equation}
\epsilon_{\rm trg}=1-P_S^{(0)}P_L^{(1)}
\end{equation}
The trigger efficiency $\epsilon_{\rm trg}$ is given in 
\Tab\ref{trgtab} for \pippim\ and \piopio\ events, and for the three different values of \ecr. 
The maximal variations in  $\epsilon_{\rm trg}$ during data taking are 0.5\% and 
0.1\%, respectively,
and are due to variations in the energy threshold of the calorimeter trigger 
(related to small changes in the gain of the calorimeter photomultipliers).
The systematic error is evaluated using MC events as the difference between the result given by the above method 
and the MC truth. It is  $0.25\!\times\!10^{-3}$ for \pippim\ events with $\ecr\!=\!300$\MeV, and 
negligible for \piopio.

The contribution of accidental clusters to the trigger gives an
additional systematic error. This is important only for the \pippim\ channel,
for which the trigger inefficiency is \ab1.3\%, as opposed to \ab0.1\% for \piopio. 
This has been studied using an independent estimate of the trigger 
efficiency for \pippim\ 
events, which is obtained by weighting MC kinematics with data-extracted trigger-sector efficiencies.
In contrast to the method for determining the trigger efficiency described above, this method 
does not include the possible contribution to the trigger from accidental clusters. 
The difference between the results from the two 
methods is $0.62\!\times\!10^{-3}$; this is taken as a further systematic error on the trigger efficiency.

The overall systematic error on the ratio of trigger efficiencies is
$0.67\!\times\!10^{-3}$ at $\ecr\!=\!300$\MeV\ (\Tab\ref{systtab}).

The cosmic-ray veto causes \ab3.5\% of the events selected with a 
\kcr\ tag to be lost.
The difference between veto efficiencies for \pippim\ and \piopio\ events is very small,
since in the majority of the rejected events the \kcr\ cluster satisfies the cosmic-ray veto by
depositing energy in two adjacent sectors of the
outermost EMC layer, and this is independent of the \ks\ decay channel. Nevertheless, veto efficiencies are 
evaluated for each channel using a subsample of 
selected events for which the cosmic-ray veto was 
present but not enforced at acquisition. 
The cosmic-ray veto efficiency
$\epsilon_{\rm CV}$ is given in \Tab\ref{trgtab} for \pippim\ and \piopio. The maximal variation
in these efficiencies during data taking is $\ab4\!\times\!10^{-3}$. 
The statistical error on the ratio of \pippim\ and \piopio\ efficiencies is
 $\ab0.2\!\times\!10^{-3}$ and is included in the
statistical error on the efficiency corrections (\Tab\ref{systtab}).

The background-rejection filter FILFO makes use of EMC cluster properties and the number of 
DC hits and is intended to eliminate machine-background or cosmic-ray events before DC reconstruction.
The ratio of FILFO efficiencies for \pippim\ and \piopio\ events
is estimated by MC to be different from unity by $\ab0.7\!\times\!10^{-3}$ 
at $\ecr\!=\!300$\MeV. 
Since FILFO is based on variables with distributions depending on the run conditions, 
such as the number of DC hits and the fraction of DC hits in the innermost DC layers,
the reliability of this prediction has been checked by studying a data subsample 
for which the FILFO decision is registered but not enforced during reconstruction.
The ratio of FILFO efficiencies for \pippim\ and \piopio\ events in data is found to be different from unity
by less than $10^{-4}$ (\Tab\ref{trgtab}), and is used to correct the prediction from MC. 
The systematic error on the ratio of FILFO efficiencies
is assumed to be equal to the ratio predicted by MC, which is
$0.74\!\times\!10^{-3}$ at $\ecr\!=\!300$\MeV\ (\Tab\ref{systtab}).
\begin{table}[ht!]
  \begin{center}
\renewcommand{\arraystretch}{1.2}    
    \begin{tabular}[c]{|c|c|c|c|c|}\hline
 \multicolumn{2}{|c|}{\ecr\ value}      &  125\MeV & 200\MeV  &  300\MeV     \\ \hline 
\pippim\ & $\epsilon_{\rm trg}$      & 0.9863(1)  & 0.9867(1)  & 0.9879(2)      \\ 
         & $\epsilon_{\rm CV}$           & 0.9646(3)  & 0.9626(4)  & 0.9598(6)      \\ 
         & $\epsilon_{\rm FILFO}$    & 0.99964(2) & 0.99963(3) & 0.99944(4)     \\ \hline
\piopio\ & $\epsilon_{\rm trg}$      & 0.99948(3) & 0.99948(3) & 0.99951(4)     \\ 
         & $\epsilon_{\rm CV}$           & 0.9625(9)  & 0.959(1)   & 0.954(2)       \\ 
         & $\epsilon_{\rm FILFO}$    & 0.99956(3) & 0.99953(3) & 0.99937(5)     \\ \hline

    \end{tabular}
  \end{center}
  \caption{Values for the trigger, cosmic-ray veto, and FILFO efficiencies for \pippim\ and \piopio\ events, 
for data sample no.\,10 and minimum \kcr\ energies of 125, 200, and 300\MeV. Statistical errors on the last 
digit are
shown in parentheses.}
  \label{trgtab}
\end{table}

\section{Results}
\label{result}
\begin{figure}[htbp!]
  \begin{center}
    \resizebox{!}{7cm}{\includegraphics{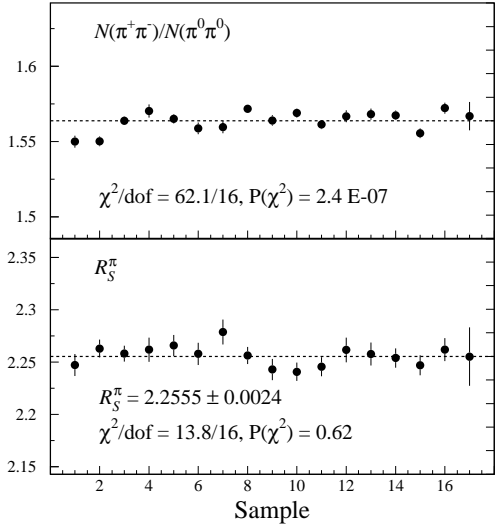}}
    \caption{Ratio $N(\pippim)/N(\piopio)$ (top) and result for \R\ (bottom)
    for $\ecr\!=\!300$\MeV, for 17 data samples. The fractional vertical range for both plots 
    is 10\%, so that 
    each tick on the right vertical axis corresponds to 1\%.
    The error bars represent the total statistical error. 
    The results of fits of $N(\pippim)/N(\piopio)$ and \R\ to constants 
    and the associated $\chi^2$ values are also shown.}
    \label{analysis300}
  \end{center}
\end{figure}

\begin{figure}[htbp!]
  \begin{center}
    \resizebox{!}{7cm}{\includegraphics{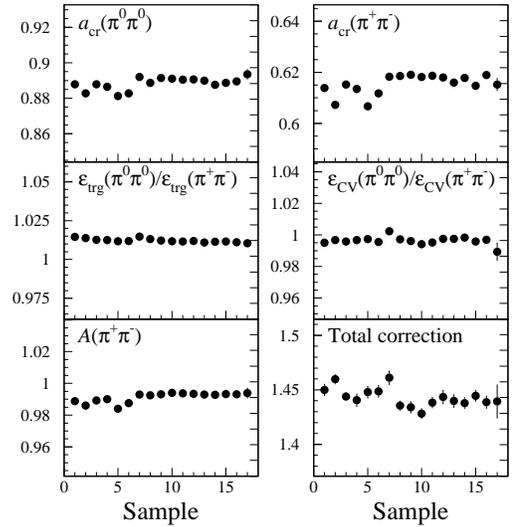}}
    \caption{Most significant efficiency corrections and total correction applied to the ratio 
    $N(\pippim)/N(\piopio)$ 
    for $\ecr\!=\!300$\MeV, for 17 data samples. The fractional vertical range for both plots 
    is 10\%, so that 
    each tick on the right vertical axis corresponds to 1\%.
    The error bars represent the total statistical error.}
    \label{effis300}
  \end{center}
\end{figure}

The ratio $N(\pippim)/N(\piopio)$ for $\ecr\!=\!300$\MeV\ 
is shown in the top panel of \Fig\ref{analysis300}. The data have been 
divided into 17 samples of comparable statistical weight; 
the first six samples correspond to data collected during 2001, samples from 7 to 16 were acquired
during 2002, and
the last sample refers to data from a dedicated scan performed by varying the center of mass energy 
by $\pm3$\MeV\ around the $\phi$ peak.
The variations observed for $N(\pippim)/N(\piopio)$ are significantly greater than the statistical fluctuations and 
are due to variations in the overall efficiencies. 
The most sizable corrections appearing in the ratio of $\piopio$ and $\pippim$ selection efficiencies
of \Eq\ref{tnpn} are shown in the first five panels of \Fig\ref{effis300}: these are 
the  acceptances for $\piopio$ and $\pippim$ events, the ratio of trigger and cosmic-ray veto efficiencies,
and the tagging-efficiency factor $A(\pippim)$. The variations observed are more pronounced for the
samples collected during 2001, for which the rates of machine background were higher and more unstable than for 2002. 
These have particularly affected the DC efficiency for the innermost layers, and therefore the $\pippim$ acceptance.

Each measurement of \R\ is obtained by correcting the
number of \pippim\ and \piopio\ events by the ratio of the selection efficiencies 
and the background contaminations (\Eq\ref{strategone}) shown in the sixth panel of \Fig\ref{effis300}.
In order to avoid statistical correlations between the event counts and
the efficiency corrections evaluated 
from data, each sample has been split into three parts on a
random basis. The first of these is used for event counting, the second for
the calculation of the tagging efficiency, and the third for the evaluation of the trigger efficiency.
The result for \R\ is shown in the bottom panel of
\Fig\ref{analysis300}; the error bars represent the total statistical error, which for most of the
samples corresponds to a fractional uncertainty of $\ab4\!\times\!10^{-3}$. 
The $\chi^2$ probability of the fit to a constant is 62\%.
All quantities entering into the measurement of \R\ are listed in Tabs.~\ref{tagtab} to~\ref{trgtab}.

The systematic errors have been evaluated for each sample separately and then averaged by weighting  
the result from each sample with the corresponding statistical error.
The various contributions to the total statistical and systematic errors are 
shown in \Tab\ref{systtab} for \kcr\ minimum energies of 125, 200, and 300\MeV; 
 the ``syst-stat'' error listed in the second row 
refers to the statistical uncertainty from all of the corrections; all of the sources of 
systematic error have been discussed in the previous sections.

The final result is obtained by choosing the value of \ecr\ which minimizes the total error. 
The best accuracy is 
obtained for a cut of 300\MeV\ (see \Tab\ref{systtab}). The result is:
 \begin{equation}
\label{presentresult}
\R\!=\!2.2555\!\pm\!0.0012_{\rm stat}\!\pm\!0.0021_{\rm syst\mbox{-}stat}\!\pm\!0.0050_{\rm syst},
 \end{equation}
where the first error is from the statistics of \pippim\ and \piopio\ events, 
the second is due to the 
statistical error in estimating all
of the corrections, and the last is the systematic 
uncertainty; again it must be emphasized that the error from event counting refers to one third of 
the total available sample.
\begin{table*}[ht!]
  \begin{center}
\renewcommand{\arraystretch}{1.2}    
    \begin{tabular}[c]{|c|c|c|c|c|}\hline
  \multicolumn{2}{|c|}{\ecr\ value}       & 125\MeV & 200\MeV & 300\MeV\\ \hline \hline
  \multicolumn{2}{|c|}{Source          }&  \multicolumn{3}{c|}{Fractional statistical error, ($10^{-3}$)} \\ \hline
  \multicolumn{2}{|c|}{Event count, ``stat''     }     & 0.34  & 0.40 & 0.54  \\ 
  \multicolumn{2}{|c|}{Efficiencies, ``syst-stat''}     & 0.55  & 0.65 & 0.93  \\ \hline 
  \multicolumn{2}{|c|}{{\bf Total statistical}}     & {\bf 0.64}  & {\bf 0.76} & {\bf 1.1}  \\ \hline\hline 
  \multicolumn{2}{|c|}{Source          }&   \multicolumn{3}{c|}{Fractional systematic error,  ($10^{-3}$)}  \\ \hline
  \pippim\ & \ks-\kl\ interference          & 0.80  & 0.80  & 0.80  \\  
           & \eto\ correction               & 2.0  & 1.8  & 1.4   \\  
           & Background                     & 0.10  & 0.10 & 0.10 \\  \hline 
  \piopio\ & Cluster counting               & 0.78 & 0.61 & 0.66  \\  
           & Wrong \tzeroev\ from \ks\      & 0.60 & 0.60 & 0.61  \\  
           & Physics background             & 0.35 & 0.14 & 0.04 \\  
           & Machine background             & 0.13 & 0.09 & 0.07  \\ \hline 
  \pippim/\piopio\ & Accidental rate \Racc\ & 0.47 & 0.48 & 0.52 \\ 
           & \tn, \pin, \pout\ evaluation           & 0.67 & 0.53  & 0.45 \\ 
           & \ek\                           & 0.39  & 0.62  & 0.44 \\ 
           & Trigger                        & 0.91  & 0.78   & 0.67 \\ 
           & FILFO                          & 0.45  & 0.46  & 0.74 \\  \hline
  \multicolumn{2}{|c|}{{\bf Total systematic}}     & {\bf 2.8}   & {\bf 2.5}   & {\bf 2.2}  \\ \hline\hline
  \multicolumn{2}{|c|}{{\bf Total   }  }     & {\bf 2.8}   & {\bf 2.6}   & {\bf 2.5}  \\ \hline 
    \end{tabular}
  \end{center}\vglue2mm
  \caption{Contributions to the statistical and systematic 
uncertainties, for minimum \kcr\ energies of 125, 200, and 300\MeV; the ``syst-stat'' error
refers to the statistical uncertainty from all corrections; all sources 
of systematic error have been discussed in section~\ref{effi}. }
  \label{systtab}
\end{table*}

Some of the corrections show variations as a function of \ecr:
the most important of these are 
the tagging efficiencies [\ek\ and $A(\pippim)$, \Tab\ref{tagtab}],
and the contamination in the \piopio\ selection ($C$, \Tab\ref{piopioacce}).
In order to check the reliability of these corrections,
the results of the analysis are compared when choosing \ecr\ values of 
125, 200, and 300\MeV.  Note that the event yield decreases by a factor of three in going from 125 to 300\MeV.
In order to avoid correlation effects in the comparison, the data set has been
split using a finer granularity, corresponding
to 94 samples, each of \ab5\Lpb\ of integrated luminosity. 
The analysis is performed using a different energy cut on each successive sample.
The $\chi^{2}$ of the three values obtained has a probability of 21\% (see \Tab\ref{tab:stabilitycrash}).
\begin{table*}[ht!]
  \begin{center}
\renewcommand{\arraystretch}{1.2}    
    \begin{tabular}[c]{|c|c|c|c|}\hline
  $\kcr$ energy cut (\MeV) & 125 & 200 & 300  \\ \hline
  \R\  &  $2.2574\pm0.0025$  &  $2.2519\pm0.0027$ & $2.2590 \pm 0.0040$ \\\hline 
$\chi^2$/dof;\kern5mm$P(\chi^2)$ & \multicolumn{3}{c|}{3.12/2;\kern5mm21\%}\\\hline
    \end{tabular}
  \end{center}\vglue2mm
\caption{Values of \R\ for \kcr\ energy cuts of 125, 200, and 300\MeV, obtained from three independent samples,
each with $1/3$ of the entire statistics. The errors
include both the ``stat'' and ``syst-stat'' contributions, as defined in the
text.
The $\chi^2$ value of a fit to a constant and its probability are also shown.}
  \label{tab:stabilitycrash}
\end{table*}

The present result (\Eq\ref{presentresult})
can be compared with the KLOE result from the analysis of the year 2000 data sample~\cite{plb_rappo},
\begin{equation}
\R ~=~ 2.236 \pm 0.003_{\rm stat} \pm 0.015_{\rm syst},
\end{equation}
where in this case the systematic error includes the statistical error from all of the corrections:
$0.015 = 0.008_{\rm syst\mbox{-}stat} \oplus 0.013_{\rm syst}$.
The error on the former result 
was dominated by the systematic uncertainty on the ratio of tagging efficiencies (0.011). The present
analysis makes use of various improvements to the evaluation of the tagging efficiencies
with respect to the analysis scheme of \Ref\onlinecite{plb_rappo}:
a larger window in \bstar\ is required and a more complete parametrization of the biases induced by 
errors in the \tzero\ evaluation has been included. As a result, 
the absolute systematic error due to the tagging efficiencies has been reduced to 0.0014.
The systematic uncertainty due to other sources have been reduced as well, from 0.0069 to 0.0048.
Nevertheless, the most significant change in the analysis with respect to that described in \Ref\onlinecite{plb_rappo}
is the improved treatment of the tag bias.
Therefore, when comparing
the two results, the statistical errors and the systematic errors on the tagging efficiencies are treated as independent
errors.
With this assumption, the two results are compatible, with a probability of $18\%$.
The two measurements can therefore be averaged. Weighting each by its independent errors and calculating
the average systematic error with the same weights gives: 
\begin{equation}
\label{resultone}
\R ~=~ 2.2549 \pm 0.0054.
\end{equation}
In \Ref\onlinecite{plbpennew}, this result is combined with the KLOE
measurements of \gammo{\DKSeIIIboth}/\gammo{\DKSpippim} to extract the dominant \ks\ BR's. 
To this end, we exploit unitarity: the sum of the BR's for the $\pi\pi$ and $\pi l\nu$ modes has been 
assumed to be equal
to one, the remaining decays accounting for less than $10^{-4}$. 
The BR of the decay \DKSmuIII\ has been evaluated from the KLOE measurement of \BR{\DKSeIII} and lepton universality.
All the results are summarized in the Appendix~\ref{appendone}.
For the $\pi\pi$ modes, we find:
\begin{equation}\eqalign{
\BR{\DKSpippim}&=(69.196\pm0.051)\% \cr
\BR{\DKSpiopio}&=(30.687\pm0.051)\% \cr}
  \label{eq:bratioresults}
\end{equation}
The KTeV collaboration, using their measurement of the ratio of BR's for the \kl, $\RL\!=\!2.283\pm0.034$, 
together with the world average for $\Re(\epsilon^{\prime}/\epsilon)$, 
$\Re(\epsilon^{\prime}/\epsilon)\!=\!(1.67\pm0.26)\!\times\!10^{-3}$,
quotes an expected value of \R~\cite{Alexopoulos:2004sx}: $\R\!=\!2.261\pm0.033$.
This is in good agreement with the present result, \Eq\ref{resultone}.

\section*{Acknowledgments}
We thank the \Dafne\ team for their efforts in maintaining low-background running 
conditions and their collaboration during all data taking. 
We want to thank our technical staff: 
G.~F.~Fortugno for his dedicated work to ensure efficient operations of 
the KLOE Computing Center; 
M.~Anelli for his continuous support to the gas system and the safety of
the detector; 
A.~Balla, M.~Gatta, G.~Corradi, and G.~Papalino for the maintenance of the
electronics;
M.~Santoni, G.~Paoluzzi, and R.~Rosellini for general support to the
detector; 
C.~Piscitelli for his help during major maintenance periods.
This work was supported in part by DOE grant DE-FG-02-97ER41027; 
by EURODAPHNE, contract FMRX-CT98-0169; 
by the German Federal Ministry of Education and Research (BMBF) contract 06-KA-957; 
by Graduiertenkolleg `H.E. Phys. and Part. Astrophys.' of Deutsche Forschungsgemeinschaft,
Contract No. GK 742; 
by INTAS, contracts 96-624, 99-37.
\appendix
\section{Evaluation of \ks\ BR's}
\label{appendone}
The main \ks\ BR's are evaluated from the measurements of \R\ and from the ratio of BR's 
$R_{e\pm}\equiv\BR{\DKSeIIIboth}/\BR{\DKSpippim}$. The measured values are~\cite{plbpennew}:
\begin{equation}\eqalign{
R_{e+} & = \SN{(\VS{\VS{5.099}{0.082_{\rm stat}}}{0.039_{\rm syst}})}{-4}\cr
R_{e-} & = \SN{(\VS{\VS{5.083}{0.073_{\rm stat}}}{0.042_{\rm syst}})}{-4}}
\label{eq:ratiosbr}
\end{equation}
The correlation between results for $R_{e+}$ and $R_{e-}$ is 13\%. 
The only remaining mode with a BR large enough to measurably affect the constraint $\sum_{f}\BR{\ks\!\rightarrow\!f}\!=\!1$ is $K_{\mu3}$; the BR's for 
all other channels sum up to $\ab10^{-5}$. 
Assuming lepton universality, 
\begin{equation}
r_{\mu e}=\frac{\BR{\DKSmuIII}}{\BR{\DKSeIII}}=\frac{1+\delta_{K}^{\mu}}{1+\delta_{K}^{e}}\frac{I_{K}^{\mu}}{I_{K}^{e}},
\end{equation}
where $\delta_{K}^{\mu,e}$ are mode-dependent long-distance radiative corrections and $I_{K}^{\mu,e}$ are 
decay phase-space integrals. Using $I_{K}^{\mu}/I_{K}^{e}=0.6622(18)$ 
from KTeV~\cite{Alexopoulos:2004sy} and $(1+\delta_{K}^{\mu})/(1+\delta_{K}^{e})=1.0058(10)$ 
from Ref.~\onlinecite{Andre:2004tk}, a value for $r_{\mu e}$ is obtained: $r_{\mu e}=0.6660(19)$.
The four main BR's of the \ks\ are evaluated from
\begin{equation}
\label{eq:bratioresultstot}
\BR{\ks\to i}=\frac{\gammo{\ks\to i}/\gammo{\DKSpippim}}{1+1/\R+(R_{e+}+R_{e-})(1+r_{\mu e})},
\end{equation}
where $i\!=\!$ \pippim, \piopio, $\pi^-e^+\nu$, $\pi^+e^-\overline{\nu}$.
The result is:
\begin{equation}\label{res:brs}\eqalign{
\BR{\DKSpippim}&=(69.196\pm0.051)\x10^{-2} \cr 
\BR{\DKSpiopio}&=(30.687\pm0.051)\x10^{-2} \cr 
\BR{\DKSeIIIeppm}&=(3.528\pm0.062)\x10^{-4}\cr 
\BR{\DKSeIIIempp}&=(3.517\pm0.058)\x10^{-4}}
\end{equation}
The correlation matrix $\langle\delta_{i}\delta_{j}\rangle/\sqrt{\langle\delta_{i}^{2}\rangle\langle\delta_{j}^{2}\rangle}$ is 
 \begin{equation}
  \begin{array}{cc}
    &
    \begin{array}{cccc}
      \pi^{+}\pi^{-}\kern3mm & 
      \pi^{0}\pi^{0}\kern3mm & 
      \kern1mm\pi^{-}e^{+}\nu\kern2mm & 
      \kern1mm\pi^{+}e^{-}\overline{\nu}\kern1mm\\
    \end{array}\\
    \begin{array}{c}
      \pi^{+}\pi^{-} \\ 
      \pi^{0}\pi^{0} \\
      \pi^{-}e^{+}\nu \\ 
      \pi^{+}e^{-}\overline{\nu}\\
    \end{array}
    &
  \left(
  \begin{array}{cccc}
   1      & 
  -0.9996 & 
   0.0254 & 
   0.0294 \\ 
  -0.9996 & 
   1      & 
  -0.0484 & 
  -0.0511 \\ 
   0.0254 & 
  -0.0484 & 
   1	 & 
   0.1320 \\ 
   0.0294 & 
  -0.0511 & 
   0.1320 & 
   1      \\
   \end{array}
   \right)
   \end{array}
 \end{equation}
\bibliographystyle{apsrev}
\bibliography{paper}

\end{document}